# Ultrafast Emergence of Ferromagnetism in Antiferromagnetic FeRh in High Magnetic Fields

I. A. Dolgikh\*, T. G. H. Blank, A. G. Buzdakov, G. Li, K. H. Prabhakara, S. K. K. Patel, R. Medapalli, E. E. Fullerton, O. V. Koplak, J. H. Mentink, K. A. Zvezdin, A. K. Zvezdin, P. C. M. Christianen & A. V. Kimel

*Radboud University, Institute for Molecules and Materials, 6525 AJ Nijmegen, The Netherlands.*

*High Field Magnet Laboratory (HFML - EMFL), Radboud University, Toernooiveld 7, 6525 ED Nijmegen, The Netherlands.*

*Center for Memory and Recording Research, University of California, San Diego, La Jolla, California 92093-0401, USA.*

**Ultrafast heating of FeRh by a femtosecond laser pulse launches a magneto-structural phase transition from an antiferromagnetic to a ferromagnetic state. Aiming to reveal the ultrafast kinetics of this transition, we studied magnetization dynamics with the help of the magneto-optical Kerr effect in a broad range of temperatures (from 4 K to 400 K) and magnetic fields (up to 25 T). Three different types of ultrafast magnetization dynamics were observed and, using a numerically calculated H-T phase diagram, the differences were explained by different initial states of FeRh corresponding to a (i) collinear antiferromagnetic, (ii) canted antiferromagnetic and (iii) ferromagnetic alignment of spins. We argue that ultrafast heating of FeRh in the canted antiferromagnetic phase launches practically the fastest possible emergence of magnetization in this material. The magnetization emerges on a time scale of 2 ps, which corresponds to the earlier reported time-scale of the structural changes during the phase transition.**

I.  INTRODUCTION

The metallic alloy FeRh stands out due to the counter-intuitive heat-induced ferromagnetism first reported in 1938 [1]. The material is antiferromagnetic at low temperatures and becomes ferromagnetic when heated above 370 K. The magnetic changes are accompanied by an expansion of the unit cell of FeRh. The nature of this heat-induced ferromagnetism in FeRh has been a subject of debates for about 60 years. The dispute of whether it is the change in the magnetic order that drives the lattice expansion or vice versa very much resembles the chicken-or-egg causality dilemma [2–6].

The very first hypothesis explaining the mechanism of the emerging ferromagnetism stated that the structural change leads to a sign change of the exchange integral and thus initiates changes of the order of the Fe-spins from antiferromagnetic to ferromagnetic [7]. However, very soon thereafter it was argued that this mechanism is inconsistent with the actual change of the lattice entropy. The latter, estimated from experimental data, appeared to be much less than the total entropy change [8]. Several computational studies proposed that the magneto-structural phase transition is entirely driven by processes in the spin system [9–13]. At the same time, one cannot ignore that according to recent calculations the changes in the lattice entropy and the total entropy during the phase transition in FeRh differ just by a factor of 3 [14] or are even comparable [3]. Several attempts have been done to detect the ultrafast kinetics during the phase transition using a femtosecond laser pulse as an ultrafast heater and tracing the laser-induced dynamics of the lattice and spins. However, even after these experiments, the dispute is far from being resolved. Refs. [15–17] claim that the changes in the spin system are faster than those in the lattice, while others [18–20] state that the lattice expands much earlier than the net magnetization emerges. In Ref. [21] it was even found that the electronic band structure changes on a subpicosecond time-scale during laser-induced phase transition far before lattice or spins. Hence, the sequence and the actual mechanism of the magnetic and structural dynamics in the magneto-structural phase transition of FeRh remain unclear. Lacking efficient means to control the speed of the dynamics of the lattice or spins, this question has become a classical chicken-or-egg causality dilemma.

In addition to these fundamental obstacles, in practice, the studies of FeRh are hampered by the co-existence of the low-temperature antiferromagnetic and the high-temperature ferromagnetic phases [22–24]. Although such a co-existence is very typical for first-order phase transitions, as in FeRh, it often leads to difficulties in interpreting the experimental data as discussed in Ref. [25].

Here we aim to understand the ultrafast magnetization dynamics during the magneto-structural phase transition in FeRh by performing time-resolved magneto-optical pump-probe measurements in an unprecedently broad range of temperatures (from 4 K to 400 K) and magnetic field strengths (from 0.125 T to 25 T). Increasing the magnetic field up to 25 T is an efficient means to control the speed of the magnetization dynamics in FeRh, while decreasing the temperature down to 4 K affects the volume ratio of the co-existing ferro- and antiferromagnetic phases.

The paper is organized as follows. Section II describes the studied sample and the experimental procedure. Section III is dedicated to the H-T phase diagram of the studied FeRh. The peculiarities of the diagram are revealed by combining a numerical approach with static measurements of magnetic and magneto-optical properties. In Section IV, we report the results of the time-resolved magneto-optical measurements of the ultrafast magnetization dynamics. In particular, it is shown that the applied magnetic field is able to accelerate the emergence of the ferromagnetic phase, but only to a certain limit, when the characteristic time is close to the characteristic time-scale of the accompanying structural changes. Moreover, the section reports that depending on temperature and applied magnetic field the femtosecond laser pulse can trigger three different types of ultrafast magnetization dynamics in FeRh. These differences are explained by three different initial states of the spin order in FeRh, corresponding to collinear antiferromagnetic, canted antiferromagnetic, and ferromagnetic phases. This interpretation is supported by simulations in Section V showing that the observed changes in the magnetization dynamics are intrinsic even to the simplistic two-spin model and thus must be a general feature of all antiferromagnets and not only FeRh. Section VI reports about the time-resolved measurements of the reflectivity, which contains information about the dynamics of the structural changes in FeRh. This section supports the hypothesis that the magnetic and structural changes during the phase transition occur simultaneously in step with each other. Finally, in section VII the conclusions are given.

## II. SAMPLE AND EXPERIMENTAL PROCEDURE

FeRh is a metallic compound with a CsCl crystal structure and a tetragonal magnetic unit cell. At low temperatures, in the absence of any external fields and at ambient pressure, the spins of the Fe atoms are aligned antiferromagnetically and have a net magnetic moment $m_{\text{Fe}} \approx 3\ \mu_B$ per atom, while the Rh atoms have no magnetic moment [26]. When increasing the temperature to around 370 K, a phase transition is triggered from the antiferromagnetic (AFM) to the ferromagnetic (FM) state. The parallel aligned magnetic moments of iron remain at $m_{\text{Fe}} \approx 3\ \mu_B$ per atom, while the Rh ions acquire an atomic magnetic moment of $m_{\text{Rh}} \approx 1\ \mu_B$ (see Fig.1b) [27,28].

These magnetic changes are accompanied by an expansion of the unit cell of about 1% [29], preserving the CsCl crystal structure [30]. The studied sample was a 40 nm thick, epitaxial Fe$_{50}$Rh$_{50}$ film deposited onto a MgO(001) single crystal substrate and then capped with a 5 nm thick Pt layer. The total volume of the FeRh(001) layer was estimated to be $2.9\times10^{-7}$ cm$^3$. More details on the structural characterization such as X-ray reflectometry, X-ray diffraction, and manufacturing procedure of the studied film are reported in Ref. [31]. Unlike bulk samples, thin films of FeRh may expand differently along different crystallographic axes [32].

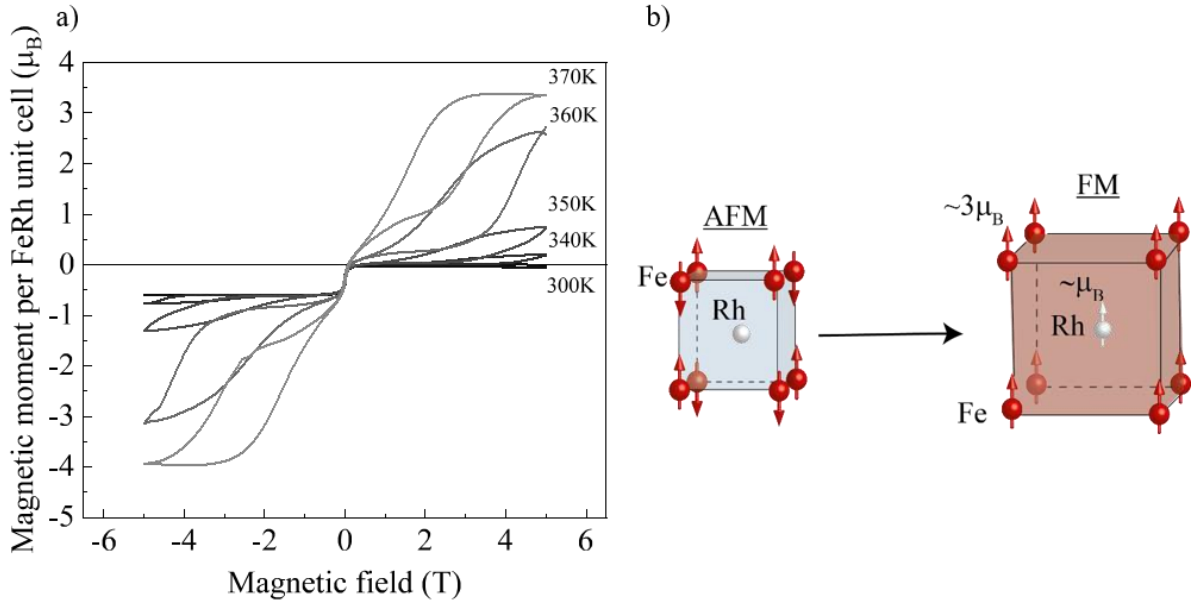

FIG. 1. (a) Magnetization as a function of magnetic field applied along the normal to the sample. Data was obtained using SQUID magnetometry of a FeRh thin film for temperatures from 300 K to 370 K. (b) Unit cell of FeRh with the schematics showing the changes during the magneto-structural phase transition from an anti-ferromagnetic (AFM) to a ferromagnetic (FM) state, accompanied by an expansion of the lattice.

In order to study the ultrafast magnetization during the magneto-structural phase transition in FeRh, we employed the principle of pump-probe measurements. An intense femtosecond laser pulse (pump) is absorbed in the metallic FeRh and thus acts as an ultrafast heater. A delayed, less intense, but equally short laser pulse (probe) detects the magnetization of the FeRh film via the polar magneto-optical Kerr effect. Detecting the magneto-optical Kerr effect as a function of the time delay between the pump and the probe pulses measures the heat-induced magnetization dynamics with subpicosecond temporal resolution. It is conventionally accepted that for metals, such as FeRh, the magneto-optical Kerr effect is a reliable probe of the magnetization for time delays above 0.5 ps and is able to show the dynamics of subpicosecond demagnetization [33].

We employed an experimental setup for time-resolved magneto-optical measurements at the High Field Magnetic Laboratory (HFML) in Nijmegen [34]. An amplified Ti:sapphire laser and an Optical Parametric Amplifier (OPA) were employed as ultrafast light sources. The studied sample was put in a cryostat and a Florida-Bitter magnet. The latter allowed us to apply external magnetic fields along the sample normal as strong as 37.5 T. Optical measurements were carried out in the polar geometry of the magneto-optical Kerr effect with the pump and the probe at normal incidence to the sample. The pump pulse had a fluence of 4.1 mJ/cm$^2$ at a central wavelength of 800 nm (1.55 eV). The pump beam was focused on a spot of approximately 50 µm in diameter. The pump-induced magnetization dynamics was observed with probe pulses having a central photon energy of 1.9 eV (660 nm). We estimated the duration of the optical pulses at the sample to be around 150 fs.

Pump pulses were effectively absorbed in the sample. Taking into account the diameter of the focused pump (50 µm), the pump fluence (4.1 mJ/cm$^2$), the specific heat of FeRh (0.35 J/g K [35]), and assuming that the sample absorbs 20 % of the pump radiation [36], we estimate that a single pump laser pulse heats the upper 10 nm layer of the FeRh film by about 230 K. The pump-induced polarization rotation ($\Delta\theta$) due to the polar magneto-optical Kerr effect and the reflectivity change ($\Delta R$) of the probe pulses reflected from the sample were detected with the help of a two-photodiode balanced detector.

## III. H-T PHASE DIAGRAM OF THE AFM-FM PHASE TRANSITION IN FeRh

a) Experimental observations

It is known that applying an external magnetic field shifts the critical temperature at which the magneto-structural phase transition in FeRh occurs [9]. Figure 1 shows the magnetization curves of the studied FeRh sample. An external magnetic field was applied along the sample normal. The magnetic moment at different temperatures was recorded by a Quantum Design SQUID magnetometer in the reciprocal-sample-option (RSO) mode. The used SQUID device measures the net magnetization of the whole sample. The data shown in the figure reveals a clear jump in the net magnetization at relatively low magnetic fields (below 1 T). This jump is does not depend on the sample temperature and is usually attributed to the magnetic response of the first few FeRh layers at the interface with the MgO substrate. Due to the strain induced by the substrate, these layers cannot accommodate any volume change across the phase transition and remain in the FM phase also at lower temperatures. An applied magnetic field saturates the magnetization of these FM layers [32].

The SQUID data also shows that magnetic fields above 1 T can induce a significantly larger magnetization. The up- and down sweeps of the field reveals a hysteresis, which is associated with a field-induced first-order phase transition from an antiferromagnetic (AFM) to a ferromagnetic (FM) state. For fields up to 5 T, the phase transition is clearly seen in the range from 350 K to 370 K. At the same time, for temperatures below 370 K, 5 T it is not sufficient to saturate the magnetization of the sample. It means that even at 5 T and below 370 K the FM and AFM phases can co-exist.

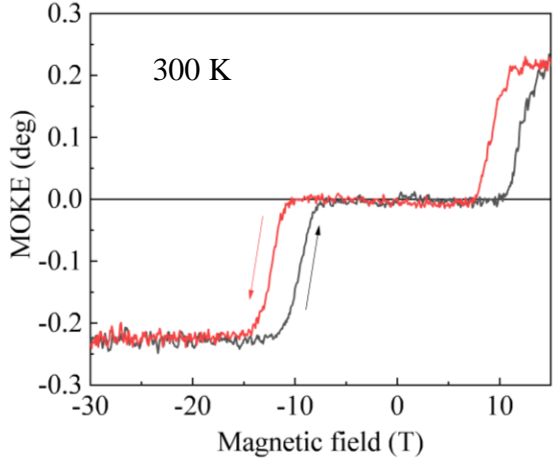

FIG. 2. Polar magneto-optical Kerr effect (MOKE) in external magnetic fields up to 30 T. The polarization rotation is due to a magneto-optical Kerr effect measured at normal incidence at 300 K. Arrows indicate the direction of the magnetic field sweep.

The magnetic moment $M$ measured in emu was recalibrated to the magnetic moment per unit cell $\mu$ in the units of the Bohr magneton $\mu_B$, using $\mu = \frac{M[emu] m_{mol}[g/mole]}{\mu_B N_A m[g]}$, where $N_A$ is Avogadro's number, $m_{mol}$ is the molar mass and $m$ is the mass of the FeRh film.

At 370 K and 3 T, the signal reaches saturation, meaning that the magnetization of the FeRh film is fully aligned by the external field along the normal to the sample, at a value close to the expected maximum ~ 4 $\mu_B$ per unit cell. Hence we assume that at 370 K a field of 3 T is sufficient to align the magnetization along the sample normal in the FM phase.

Figure 2 shows the results of static polar magneto-optical Kerr measurements at room temperature. The effect results in a polarization rotation of the reflected light, which is proportional to the normal component of the net magnetization in the sample. Light with a wavelength of 660 nm was used at normal incidence and with a magnetic field applied along the sample normal. The light beam was focused to a spot with a diameter of 50 µm. To extract the contribution of the FeRh film we have subtracted the background line. This line has been fitted to the slope of the MOKE signal after the phase transition is completed. It is seen that at 300 K the magneto-optical signal is not sensitive to the FM phase at the interface with the substrate. The difference between the MOKE and SQUID magnetometry results is due to the fact that the magneto-optical Kerr effect probes the magnetization in a thin surface layer defined by the penetration depth of light which is about 10 nm, whereas the SQUID probes the magnetization in the entire sample. The difference in sensitivities between the polar MOKE and SQUID techniques becomes especially important in the case of thin film samples, where the magnetic properties are expected to be inhomogeneous over the film thickness [32]. Clearly, magnetic fields above 1 T induce a step-like change in the MOKE signal, similar to the SQUID measurements, including the presence of hysteresis, resulting from the magneto-structural first-order AFM-to-FM phase transition. To extract the magnetic fields at which the magneto-structural phase transition takes place, we took the first derivatives of the magnetization curves, deduced the average magnetic field between the fields corresponding AFM-to-FM and FM-to-AFM transitions. We added these data points to the temperature dependence of the critical field $H_C$ (see Fig.3), obtained from earlier published data measured on FeRh samples with the same or similar compositions [9,30,37]. Although the transition is expected to be accompanied by a temperature hysteresis [38], only one critical temperature has been reported in Refs [9,30] from pulsed magnetic field experiments up to 28 T.

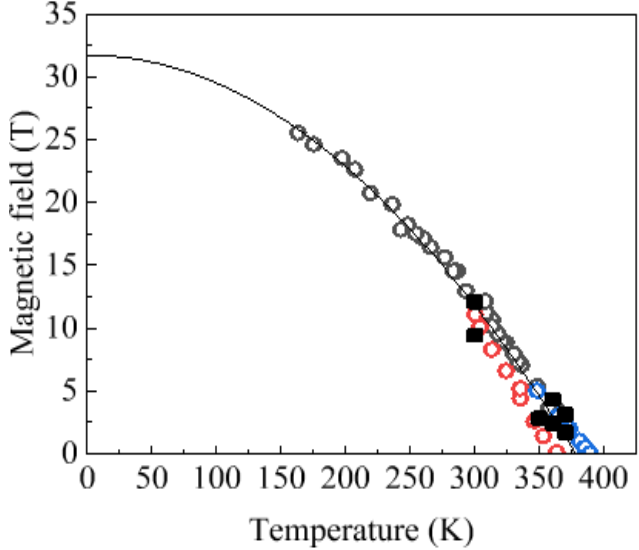

FIG.3. Critical magnetic field $H_C$ that triggers the magneto-structural phase transition from the antiferromagnetic to ferromagnetic state in $Fe_{50}Rh_{50}$. The open circles illustrate previously obtained results of the critical magnetic field of the transition for samples of similar composition: black [9], red [30], and blue [37]. The black squares represent the SQUID and MOKE data for our sample as explained in the text. The black solid line describes the fit by the empirical law from [9] $H_C(T) = H_C(0)\left(1 - \left(\frac{T}{T_c}\right)^2\right)$, where $T_c = 378 \pm 2$ K and $H_c(0) = 31.7 \pm 0.6$ T.

The SQUID magnetometry data from Ref. [37] is included by plotting the averaged value of the two temperatures of the phase transition obtained from the heating and cooling curves. The black solid line represents a fit to all literature data to the expression $H_C(T) = H_C(0)\left(1 - \left(\frac{T}{T_c}\right)^2\right)$ as proposed in Ref. [9], using $T_c = 378 \pm 2$ K and $H_c(0) = 31.7 \pm 0.6$ T. Our SQUID magnetometry and MOKE results are consistent with the literature data, so we conclude that our MOKE measurements reliably probe the phase transition from the AFM to the FM state in FeRh.

b) Theoretical modelling

Even though FeRh *H-T* phase diagrams such as the one in Fig. 3 were discussed several times before [9,12,30,35,37–44], it is clear that this diagram is not complete. If a magnetic field is applied perpendicular to the antiferromagnetic spins, the field cants spins over an angle defined by the ratio of the applied magnetic field to the effective field of the exchange interaction between the spins [45]. The canting angle increases with increasing applied field. If the magnetic field is applied parallel to the antiferromagnetically coupled spins, these spins remain insensitive to the field until the so-called spin flop field $H_{\text{sf}}$ is reached. In order to estimate $H_{\text{sf}}$ for FeRh, we write the corresponding thermodynamic potential similar to the one used in Ref. [25], but upgraded with terms accounting for magnetic anisotropy and the interaction of spins with the external magnetic field:

$$F_{eq} = -\left(J^{(2)}_{\text{Fe-Fe}}(T) + \frac{\rho_J^2}{2\epsilon}\right)(\mathbf{M}_1 \cdot \mathbf{M}_2)^2 - J^{(1)}_{\text{Fe-Fe}}(T)\mathbf{M}_1 \cdot \mathbf{M}_2 - \mathbf{H}_0(\mathbf{M}_1 + \mathbf{M}_2)$$
$$+ K(T)V/2\left(\frac{(\mathbf{M}_1 \cdot \hat{\mathbf{z}})^2}{\mathbf{M}_1^2} + \frac{(\mathbf{M}_2 \cdot \hat{\mathbf{z}})^2}{\mathbf{M}_2^2}\right) \tag{1}$$

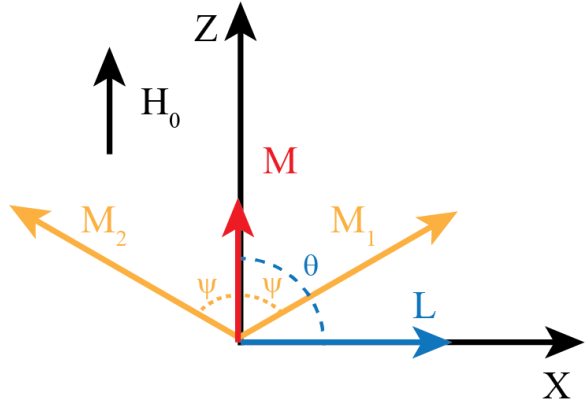

FIG. 4. Coordinate system used for the calculations of the free energy. Here, $\psi$ - the angle between $\mathbf{M_i}$ and the magnetization $\mathbf{M} = (\mathbf{M_1} + \mathbf{M_2})/2$; $\theta$ is the angle of the antiferromagnetic vector $\mathbf{L} = (\mathbf{M_1} - \mathbf{M_2})/2$ with respect to the z-axis. $\mathbf{H_0}$ denotes the direction of the external magnetic field.

Here $\mathbf{M_1}$, $\mathbf{M_2}$ ($|\mathbf{M_1}| = |\mathbf{M_2}| = M \approx 1.5$ $\mu_B$ per unit cell) are the magnetizations of the two iron sublattices with opposite spin in the AFM phase. In the first term $\rho_J$ is a lattice-dependent exchange constant as proposed in Ref. [7], $\epsilon$ is the stiffness constant [7,30,46] and $J^{(2)}_{Fe-Fe}(T)$ is an effective four-spin iron-iron exchange constant as introduced in Refs. [43,44]. The second term is given by Heisenberg exchange with $J^{(1)}_{Fe-Fe}(T)$ the temperature-dependent isotropic Heisenberg iron-iron exchange constant. The third term defines the interaction of $\mathbf{M_1}$ and $\mathbf{M_2}$ with the magnetic field $\mathbf{H_0}$ applied along the z-axis (see Fig. 4). The last term describes the magnetic anisotropy, with $K(T)$ is the constant of magnetic anisotropy, which is, in principle, temperature dependent. In this case, the anisotropy is defined such that it favours alignment of $\mathbf{M_1}$ and $\mathbf{M_2}$ along the z-axis. $V$ is the unit cell volume.

Introducing the angle $\psi$ as defined in Fig.4, we describe the orientations of the magnetizations of the Fe sublattices ($\mathbf{M_1}$ and $\mathbf{M_2}$) with respect to the magnetization $\mathbf{M} = (\mathbf{M_1} + \mathbf{M_2})/2$. $\theta$ is the angle formed by the antiferromagnetic vector $\mathbf{L} = (\mathbf{M_1} - \mathbf{M_2})/2$ and the z-axis.

In the case of the "easy-axis" of magnetic anisotropy considered here, one gets $KV < 0$, antiferromagnetic coupling favored by the two-spin exchange interaction $J^{(1)}_{Fe-Fe}(T) < 0$, and ferromagnetic coupling favored by the second exchange term $J_2 = J^{(2)}_{Fe-Fe} + \frac{\rho_J^2}{2\epsilon} > 0$.

Minimization of this total free energy with respect to the angles $\psi$ and $\theta$ gives the conditions for the equilibrium states. The collinear antiferromagnetic phase corresponds to ($\psi = \pi/2, \theta = 0$) and the ferromagnetic phase to ($\psi = 0, \theta = \pi/2$). In the canted antiferromagnetic phase in an external magnetic field equal to the spin-flop field $H_{sf}$ one can just expect $\psi = \psi_{sf}, \theta = \pi/2$, where $\cos \psi_{sf} \ll 1$. Hence neglecting terms of higher order, such as those $\sim \cos^4 \psi$, at the spin-flop field the free energy in the canted antiferromagnetic phase can be simplified to:

$$F_{sf}(\psi_{sf}) \approx \left(4J_2 M^4 - 2J^{(1)}_{Fe-Fe}M^2 + KV\right)\cos^2\psi_{sf} - 2H_{sf}M\cos\psi_{sf} - J_2 M^4 \qquad (2)$$
$$+ J^{(1)}_{Fe-Fe}M^2.$$

As the thermodynamic equilibrium implies $\frac{\partial F_{sf}(\psi_{sf})}{\partial \psi_{sf}} = 0$, it is easy to find the spin-flop angle:

$$\cos\psi_{sf} = \frac{H_{sf}}{2H_E - H_A}, \qquad (3)$$

where $J_{eff} = J^{(1)}_{Fe-Fe} - 2\left(J^{(2)}_{Fe-Fe} + \frac{\rho_J^2}{2\epsilon}\right)M^2$, $H_E \equiv -J_{eff}M$, and the anisotropy field $H_A = -\frac{KV}{M}$.

The free energy in the collinear antiferromagnetic state is equal to

$$F_{col} = -J_2 M^4 + J^{(1)}_{Fe-Fe}M^2 + KV. \qquad (4)$$

The spin-flop field $H_{sf}$ is defined as the field for which $F_{sf}(\psi_{sf}) = F_{col}$. Thus, comparing the free energies of the canted and the collinear antiferromagnetic phases, given by Eq.(2) and Eq.(4), one finds:

$$H_{sf} = \sqrt{H_A/2(2H_E - H_A)} \approx \sqrt{H_A H_E} \qquad (5)$$

Using Eq.(3) and Eq.(5) we can estimate $H_{sf}$ and $\psi_{sf}$ assuming $J_{eff}$ of the order reported in Ref. [11]. Performing numerical minimization of the thermodynamic potential given by Eq.(1), we can also analyze the whole *H-T* phase diagram. We assumed that $J_{eff}(T)$ is linear with temperature and changes sign at the transition temperature, and the uniaxial anisotropy is out-of-plane. We note, however, that there is no experimental data on the type and strength of the magnetic anisotropy in the antiferromagnetic phase of FeRh. However, thin magnetic films with zero magnetization, such as ferrimagnets at the compensation temperature, favour out-of-plane magnetic anisotropy due to the absence of demagnetizing fields and a dominant surface anisotropy contribution, resulting from the breaking of the inversion symmetry at the interfaces of the film. Such out-of-plane anisotropy in the AFM phase of FeRh is also predicted by computational studies in Refs. [47,48]. Hence we assumed that the value for the constant of magnetic anisotropy $K$ is of the order of those proposed in Refs. [48–50]. To further simplify the model, we assume that the magnetic anisotropy is temperature independent. The actual parameters used in the thermodynamic potential are given by: $J_{Fe-Fe}^{(1)} M^2 = 0.46 \cdot 10^{-14} (erg)$, $J_2 M^4 = 0.23 \cdot 10^{-14} (erg)$, $K = 5 \cdot 10^6 (erg/cc)$ and the resulting *H-T* phase diagram is shown in FIG 5a. We see that the model reproduces all three expected phases: collinear antiferromagnetic (light blue area), canted antiferromagnetic (dark blue area), and ferromagnetic (brown area). The calculated critical fields of the phase transition to the FM state (transition from blue to brown areas) are remarkably close to those obtained in our experiment, although in our model for most applied fields this transition to the FM phase starts from the canted AFM phase, rather than from the collinear phase. Only in the low field region the model predicts a direct transition from the collinear AFM phase to the FM phase. We note that the spin-flop transition (from the collinear AFM phase to the canted AFM phase) has not yet been reported for FeRh, but we envisage that its experimental observation must be seriously hampered by the relatively small angle of the spin canting just after the spin-flop $\psi_{sf} = 5° - 10°$, estimated using our model.

One can argue that there is no real proof of the fact that in our experiment FeRh has a transition between collinear and canted AFM phases. However, at the same time, there is also no experimental proof that the magnetic anisotropy of antiferromagnetic FeRh is in-plane. In this case, the fields applied in our experiment would tilt the spins even easier, but this canting has also never been reported experimentally (see also Fig. 2).

Hence the question about the peculiarities of this phase diagram justifies further experimental studies. To this end, in the following section, we report on our experimental study of the ultrafast magnetization dynamics, the results of which agree with the hypothesis that the AFM phase of FeRh has out-of-plane magnetic anisotropy and that a sufficiently high out-of-plane magnetic field triggers a spin-flop phase transition between the collinear and canted AFM phases.

## IV. EXPERIMENTAL STUDY OF ULTRAFAST MAGNETIZATION DYNAMICS

The horizontal black arrows in Figure 5a show that by properly choosing the values of the applied magnetic field $H$ and temperature $T$, one should be able to start an ultrafast heating experiment from different phases. For instance, at $T = 200$ K and $H = 5$ T, heating over a range of 230 K should cause a phase transition from the collinear AFM to the FM phase (denoted by the route type 1 arrow). At $T = 200$ K and $H = 10$ T, the same heating should trigger a transition from the canted AF to the FM phases (route type 2). Finally, at 200 K and the highest field $H = 25$ T, the heating would partially destroy the ferromagnetic order leading to demagnetization (route type 3).

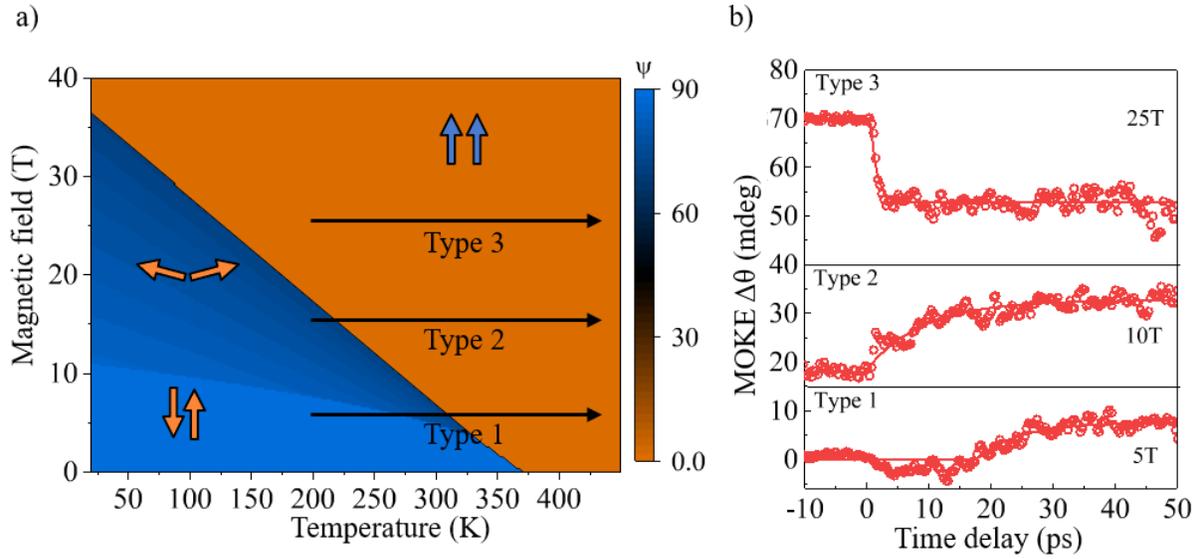

FIG. 5. (a) Numerically calculated *H-T* phase diagram using the thermodynamic potential described in the text: light blue region – collinear AFM phase, dark blue region – canted AFM phase, brown region – FM phase. The horizontal black arrows schematically illustrate the three different types of pathways into the FM phase upon ultrafast laser heating at 200 K for 230 K, depending whether the initial phase is collinear AFM (type 1), canted AFM (type 2) or FM (type 3) (b) The laser-induced change in the polarization rotation Δθ (red circles) shows the polar magneto-optical Kerr effect at applied magnetic fields of 5 T, 10 T, and 25 T corresponding to routes of types 1, 2, and 3, respectively. All the curves were taken at T = 200 K. The solid red lines are guides for the eye obtained by fitting function $\Delta\theta(t) = B_M\left(1 - e^{-\frac{t-\Delta\tau}{\tau_M}}\right)$ to the data. For route type 1, a latency appears in the dynamics, similar to Ref. [25], which is estimated to be Δτ = 17.0 ± 0.6 ps. There is no latency as well as $B_M > 0$ for route type 2 and $B_M < 0$ for route type 3.

Interestingly, if the heating is ultrafast and realized with the help of fs-laser pulses [15,16], each of these three routes must have distinctly different kinetics. Indeed, the dynamics of spin **S** in the effective exchange field of its neighbouring spins **H**$_E$ must obey the fundamental law of conservation of angular momentum $\frac{d\mathbf{S}}{dt} = -\gamma[\mathbf{S} \times \mathbf{H}_E]$, where $\gamma$ is the gyromagnetic ratio.

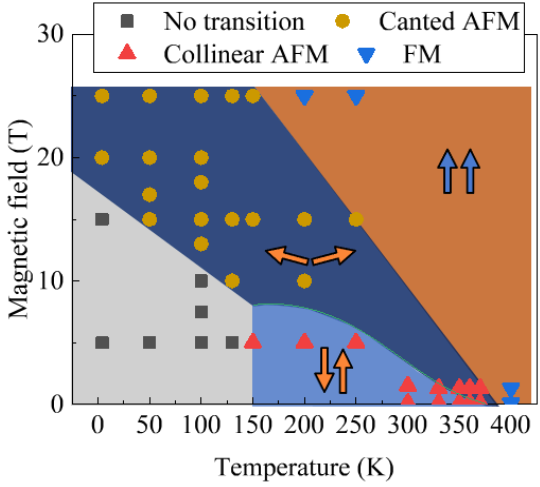

FIG. 6. Phase diagram deduced from laser-induced dynamics of the magneto-optical Kerr effect represented in Fig. S3. Every time-resolved trace was assigned to a certain route type of ultrafast dynamics depending on the sign of the amplitude $B_M$ and the presence of latency ($\Delta\tau$). In the blue area (and $\Delta\tau>0$) laser excitation triggers dynamics along Route 1 (red triangles). In the dark blue area ($B_M > 0$ and $\Delta\tau=0$), the observed dynamics is assigned to Route 2 (brown circles). In the brown area ($B_M < 0$ and $\Delta\tau=0$), the dynamics is assigned to Route 3 (blue triangles). In the grey area, the dynamics showed too low amplitudes, which could not be reliably assigned to any of these three types (black squares).

When the system is in the collinear antiferromagnetic state, an instantaneous change of the sign of the effective exchange interaction $\mathbf{H}_E$ does not immediately launch spin dynamics. Since the spins are collinear, the torque acting on the spin is zero $[\mathbf{S} \times \mathbf{H}_E] = 0$, and starting the spin dynamics requires an additional trigger leading to a latency $\Delta\tau$ as shown in Ref. [25]. When the antiferromagnet is in a canted state $[\mathbf{S} \times \mathbf{H}_E] \neq 0$ and a change of the exchange interaction will instantaneously launch the spin dynamics. Therefore, it is clear that the kinetics of the phase transition to the FM phase from the collinear and from the canted AFM phases must be substantially different. Finally, if FeRh is already in a FM phase, ultrafast heating with the help of a femtosecond laser pulse results in ultrafast demagnetization [31]. Aiming to reveal how the ultrafast magnetization dynamics actually change upon changing the initial phase, we perform a time-resolved pump-probe magneto-optical study of FeRh in fields up to 25 T at several base temperatures.

The resulting laser-induced dynamics for routes 1, 2, and 3 at $T = 200$ K in fields 5 T, 10 T, and 25 T are shown in Fig. 5(b) (red circles). At $H = 5$ T, the MOKE signal exhibits a profound latency $\Delta\tau$ similar to Ref. [25], until the MOKE signal starts to rise (similar behaviour can be observed at higher temperatures, see Fig. S1 of Supplementary materials). The polarization rotation can be fitted with the function $\Delta\theta(t) = B_M \left(1 - e^{-\frac{t-\Delta\tau}{\tau_M}}\right)$ (solid red curves in Fig 5(b). Here $B_M$ and $\tau_M$ are the amplitude and the characteristic time corresponding to the laser-induced magnetic changes. $\Delta\tau$ is the latency, similar to the one reported in Ref. [25]. From the fit of the data obtained at $H = 5$ T, we find that $B_M > 0$ and $\Delta\tau = 17.0 \pm 0.6$ ps. At $H = 10$ T, the observed dynamics is substantially different and no latency is seen ($B_M > 0$, $\Delta\tau = 0$). The magnetization starts to grow immediately after the pump excitation. Further increase of the magnetic field strength up to $H = 25$ T substantially changes the dynamics. The amplitude changes sign ($B_M < 0$) and the dynamics acquires characteristic features of ultrafast demagnetization ($\Delta\tau = 0$). These three types of dynamics can be attributed to the three types of routes shown in Fig. 5(a). To illustrate the behaviour within each of the three routes, we measured the laser-induced dynamics at various magnetic fields and temperatures. Each of the observed transients has been assigned to one of the three types shown in Fig.5 (b): type 1 ($B_M > 0$, $\Delta\tau \neq 0$), type 2 ($B_M > 0$, $\Delta\tau = 0$) and type 3 ($B_M < 0$, $\Delta\tau = 0$). To simplify the assignment procedure, we neglect changes of $\Delta\theta$ less than 10% of the maximum demagnetization signal. For instance, the dynamics obtained at 200 K and 5 T, as shown in Fig 5 (b), has been attributed to type 1, even though between 0 ps and 18 ps the signal is strictly speaking not zero. The results of the classification are shown in Fig. 6, where one can clearly distinguish three different regions for those cases when the dynamics exhibited a sufficiently high amplitude The resulting regions with routes of type 1, 2, and 3 are in close agreement with the regimes where we expected to find antiferromagnetic collinear, antiferromagnetic canted, and ferromagnetic starting phases (see Fig. 5(a)), respectively.

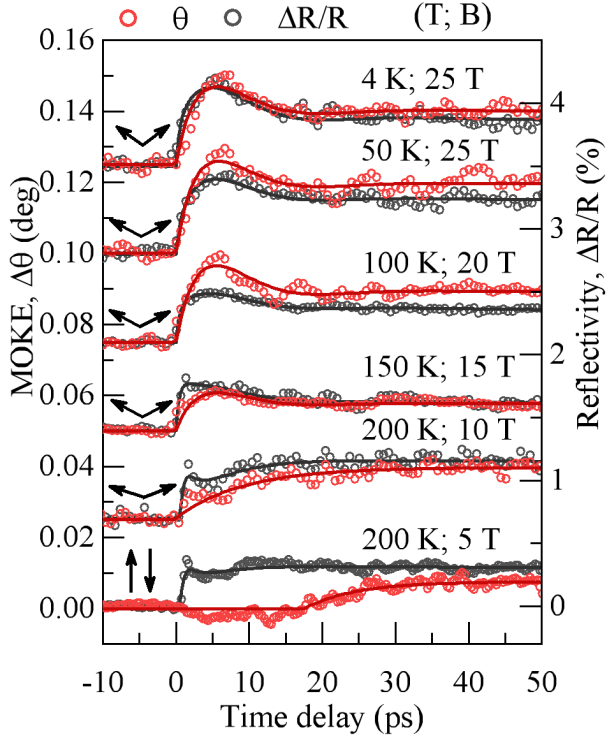

FIG. 7. Ultrafast kinetics of the magneto-structural transition. The polarization rotation due to the magneto-optical Kerr effect (red) and reflectivity change ΔR/R (black) at various temperatures and magnetic fields. Open circles represent the experimental data and the solid lines are their respective fits. The curves are plotted with an offset along the y-axis.

Hence, Figs. 5(a) and 6 support our hypothesis that by varying the external magnetic field one can substantially modify the dynamics triggered by ultrafast heating. Note that at low fields and low temperatures, transients could not be assigned to one of the three types (black squares in Fig 6). We believe that in this case, the pump pulse does not provide enough heat to reach the FM phase.

In order to reveal the fastest possible phase transition from an antiferromagnetic to a ferromagnetic phase, one should focus on the dynamics without latency and thus explore route type 2 in ever-higher magnetic fields.

Figure 7 shows a selection of the typical time traces of the MOKE signal (open red symbols) measured at temperatures down to 4 K and in magnetic fields up to 25 T. While the MOKE signal at 5 T is characterized by a latency, at higher fields, the dynamics changes dramatically and the latency disappears (see also dynamics at fixed T= 100 K Fig. S2). Trying to fit the data obtained for fields above 10 T with the function $\Delta\theta(t) = B_\text{M}\left(1 - e^{-\frac{x}{\tau_M}}\right)$ we failed to reproduce the experimental trend, which implies that the real dynamics is more complex. We note that at low temperatures (< 150 K in Fig. 7) a strongly damped oscillation can be seen in the data. In order to account for this transient feature, in the fitting function we added a part corresponding to the damped harmonic oscillator: $\Delta\theta(t) = C_M \cdot sin(2\pi F\, t\,)e^{-\gamma t} + B_M\left(1 - e^{-\frac{x}{\tau_M}}\right)$, where $C_M$ is the amplitude of the transient. We assumed that the damping is very strong $\gamma = 0.15\frac{1}{ps}$ and fitting this function to the data gives $F$ = 33±9 GHz (full details of this fitting procedure and the defined parameters can be found in the Supplementary information).

Most importantly, the fit reveals that the characteristic rise time $\tau_M$ hardly depends on the strength of the magnetic fields above 15 T (see Supplementary information). The average rise time for fields above 15 T gives $\tau_M = 2.0 \pm 0.4$ ps. Hence, for the fits shown in Fig.7 and in the rest of the paper, we fixed the time $\tau_M = 2$ ps and fitted only the amplitudes.

### V.    MODELLING ULTRAFAST MAGNETIZATION DYNAMICS

While modelling ultrafast magnetization dynamics in FeRh remains to be a challenge, it is possible to show that a sudden change of the exchange interaction in collinear and canted antiferromagnetic state triggers substantially different dynamics even in a simple case of two antiferromagnetically coupled macrospins. To this end, we consider a simplified free energy F that accounts for the exchange interactions between two macrospins $\mathbf{M}_1$ and $\mathbf{M}_2$. The latter mimic the two antiferromagnetically coupled Fe-sublattices in FeRh. The free energy also accounts for the interaction of the applied magnetic field $\mathbf{H}_0$ with the macrospins as well as assumes that the field is applied along the easy axis of magnetic anisotropy (z-axis) with uniaxial anisotropy constant $K$:

$$F = f(\mathbf{M}_1^2) + f(\mathbf{M}_2^2) - J\mathbf{M}_1 \cdot \mathbf{M}_2 - \mathbf{H}_0 \cdot (\mathbf{M}_1 + \mathbf{M}_2) + K\left(\frac{(\mathbf{M}_1 \cdot \hat{\mathbf{z}})^2}{\mathbf{M}_1^2} + \frac{(\mathbf{M}_2 \cdot \hat{\mathbf{z}})^2}{\mathbf{M}_2^2}\right) \quad (6)$$

here $f(\mathbf{M}_1^2)$ and $f(\mathbf{M}_2^2)$ are the non-equilibrium exchange energy of sublattices, as defined Ref. [51], and $J$ is the exchange constant that favours antiferromagnetic coupling between the macrospins in the ground state.

The dynamics of the macrospins is described by the set of equations describing both longitudinal and transverse spin dynamics [52,53]:

$$\frac{\hbar}{\gamma}\frac{d\mathbf{M}_1}{dt} = \mathbf{M}_1 \times \mathbf{H}_1 + \lambda_\mathrm{r}\mathbf{H}_1 + \lambda_\mathrm{e}(\mathbf{H}_1 - \mathbf{H}_2), \quad (7)$$

$$\frac{\hbar}{\gamma}\frac{d\mathbf{M}_2}{dt} = \mathbf{M}_2 \times \mathbf{H}_2 + \lambda_\mathrm{r}\mathbf{H}_2 + \lambda_\mathrm{e}(\mathbf{H}_2 - \mathbf{H}_1), \quad (8)$$

where the effective fields $\mathbf{H}_i \equiv -\frac{\delta F}{\delta \mathbf{M}_i}$ are derived from the free energy $F$ given by Eq. 7 [51]:

$$\mathbf{H}_{1,2} = -A(\mathbf{M}_{1,2}^2 - \mathbf{M}_0^2)\mathbf{M}_{1,2} + J\mathbf{M}_{2,1} + \mathbf{H}_0 - 2K\frac{\mathbf{M}_{1,2} \cdot \hat{\mathbf{z}}}{\mathbf{M}_{1,2}^2}. \quad (9)$$

Here $A$ is a non-equilibrium exchange constant [51] and $M_0$ is the saturation magnetization of the iron sublattices. The relativistic constant $\lambda_\mathrm{r}$ describes angular momentum transfer to the environment, $\lambda_\varepsilon$ allows for exchange-mediated transfer between the sublattices, $\gamma$ is the gyromagnetic ratio of iron.

In the model, the action of the laser pulse is assumed to change the sign of the exchange constant $J \rightarrow -J$ within 30 fs. Parameters used for modelling are $J = -24.8$ meV, $K = 9.75 \times 10^{-5}$ J, $A = 4$ J, $M_0 = 0.1$, $\lambda_\varepsilon = 0.05$, $\lambda_\mathrm{r} = 0.2\, \lambda_\varepsilon$ [11,15,19] The magnetization of the macrospins, as well as the relaxation parameters, are dimensionless and, the former is set to unity. The model gives for the spin-flop field $\mu_0 H_{sf} = 6$ T. The results of the simulations of the magnetization dynamics are depicted in Fig. 8 for the cases of low magnetic field $\mu_0 H_0 = 0.6$ T and the field above the spin-flop field $\mu_0 H_0 = 9$ T, where the macrospins are initially in a canted antiferromagnetic state. It is clearly shown that a high magnetic field above the spin-flop field accelerates the dynamics in agreement with our experimental findings.

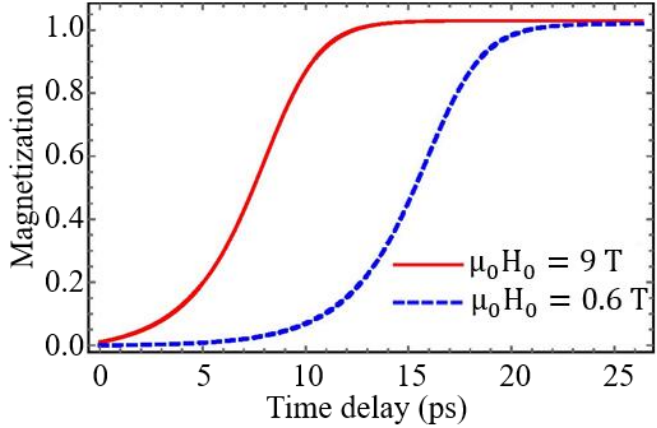

FIG. 8. Simulated magnetization dynamics of a two-sublattice antiferromagnet with ultrafast exchange-inversion occurring at zero-time delay. Using the set of parameters as explained in the text, the system has a spin-flop transition at $\mu_0 H_{sf} = 6$ T. When starting far below this field (0.6 T), the modelled dynamics shows a clear latency period in the growth of the magnetization along the applied field. This is qualitatively different from the situation in high fields ($\mu_0 H_{sf} = 9$ T), where the system initially resides in the canted antiferromagnetic state.

## VI. TIME-RESOLVED REFLECTIVITY MEASUREMENTS

Almost 20 years ago it was suggested that ultrafast lattice relaxation resulting from a magneto-structural phase transition in FeRh can be monitored by measuring time-resolved reflectivity changes (Ref. [15]). The reported time-scale for the observed lattice expansion was 5-10 ps. Lattice expansion upon a phase transition from an AFM to FM state in FeRh within the first 6 ps was confirmed by time-resolved X-ray diffractometry [20]. In our experiments, time-resolved reflectivity changes were measured simultaneously with the MOKE dynamics. As explained in Ref. [15], the obtained transients (black symbols in FIG. 7) contain contributions from the electronic part and the lattice part. For instance, the rapid increase and relaxation of the reflectivity signal in the first 2 ps after ultrafast laser excitation are conventionally assigned to be due to ultrafast heating and cooling of free electrons. This contribution to the signal is also present in the absence of the AFM- FM phase transition, so we estimated this part of the signal by analysing the reflectivity at low magnetic fields far away from the AFM – FM transition i.e. at the experimental conditions when no signatures the laser-induced phase transition were visible in the MOKE signal. Subsequently, the fit of this dynamical signal was taken as a fixed contribution in all other fits. The growth on a long timescale, which appears close to the AFM - FM phase transition, is fitted with a similar expression as for the magnetization dynamics: $\frac{\Delta R}{R}(t) = C_R \cdot \sin\left(2\pi F\, t - \frac{\pi}{2}\right) e^{-\gamma t} + A_L \left(1 - e^{-\frac{t}{\tau_L}}\right)$, where $A_L$ and $\tau_L$ are the amplitude and the characteristic time presumably corresponding to the laser-induced lattice expansion, $C_R$ is the amplitude of the damped harmonic oscillator, $F$ and $\gamma$ are its frequency and the decay rate. The fits are shown in the figure by the black solid curves. Also for the reflectivity data, we found that the rise time $\tau_L$ does not depend on the magnetic field strength (see Supporting information) with an average value of $\tau_L = 2.7 \pm 0.9$ ps. As a result, for all fits in the rest of the paper, we fixed the time $\tau_L = 2$ ps and fitted only the amplitudes.

If the observed reflectivity and the MOKE dynamics correspond to the magneto-structural phase transition, the amplitudes $A_L$ and $B_M$ must be proportional to the probability of the nucleation of the FM phase. The latter, in accordance with the Arrhenius equation, must be proportional to $\exp(-\frac{E_a}{kT})$, where $k$ is the Boltzmann constant and $E_a$ is the activation energy for the nucleus [54]. In the case of FeRh, $E_a$ is a function of the applied magnetic field. Plotting $A_L$ and $B_M$ as a 3D graph (FIG. 9) we see qualitative agreement with the Arrhenius equation - the amplitudes increase with increasing either the temperature or the magnetic field strength.

This trend is observed as long as most of the probed FeRh volume is in the AFM state. Approaching the critical magnetic field and temperature, co-existing FM domains will start to dominate the signal and the trend changes completely. All these observations point out that the amplitudes $A_L$ and $B_M$ serve as a measure of the volume, which undergoes structural and magnetic changes at the phase transition from the AFM to the FM states (comparison for higher temperatures as well is shown in Supplementary Fig. S4).

If one assumes the transient reflectivity probes structural dynamics of FeRh, the rise time $\tau_L = 2$ ps is expected to be close to the time of expansion of the studied sample area, which can be roughly estimated as $\tau_L \approx \frac{d_{\text{depth}}}{v}$, where $v \approx 4.8 \cdot 10^3$ m/s is the speed of sound approximately predicted in Ref. [55] and $d_{\text{depth}}$ is the penetration depth of the probe light in the film at the probe wavelength. According to Ref. [20], the penetration depth for pump and probe pulses in our experiment is about 15 nm. If one takes into account the thickness of the capping Pt layer, we obtain $d_{\text{depth}} \approx 10$ nm, leading to $\tau_L \approx \frac{d_{\text{depth}}}{v} \approx 2.1$ ps, which is indeed close to the value found experimentally. It is unclear, however, if the time required for the expansion should also account for the thickness of the Pt capping layer. For instance, one can argue that the laser pulse launches an expansion wave at the Pt-vacuum interface and it will take more than 1 ps for this expansion wave to arrive at the Pt-FeRh interface. Only afterwards reflectivity data will contain information about the lattice dynamics of FeRh. We argue, however, that our experiments do not support this hypothesis. For instance, at high temperatures, when femtosecond laser pulses launch ultrafast demagnetization of the FM-phase, or in high magnetic fields, when the fastest possible magnetization dynamics in the AFM phase is observed, the signals of transient reflectivity and MOKE start their rises practically simultaneously. As the MOKE, in contrast to reflectivity, is dominated by the magnetization of Fe-spins, based on these observations we tend to conclude that femtosecond laser pulse launches both reflectivity and MOKE dynamics upon reaching FeRh/Pt interface. The fitting and averaging procedures gave very close values for the characteristic rise times for the laser-induced MOKE and reflectivity signals: $\tau_M = 2.0 \pm 0.4$ ps, $\tau_L = 2.7 \pm 0.9$ ps. In any case, even from the experimental raw data, it is directly evident that $\tau_M \approx \tau_L$. In addition, we observed a magnetic field independent (damped) oscillatory signal in both MOKE and reflectivity channels. Similar oscillations were observed in XRD experiments and interpreted as a result of laser-induced strain wave [56]. Our observation of such an oscillation in the MOKE signal further underlines mutual correlations between the magnetic and structural dynamics in FeRh.

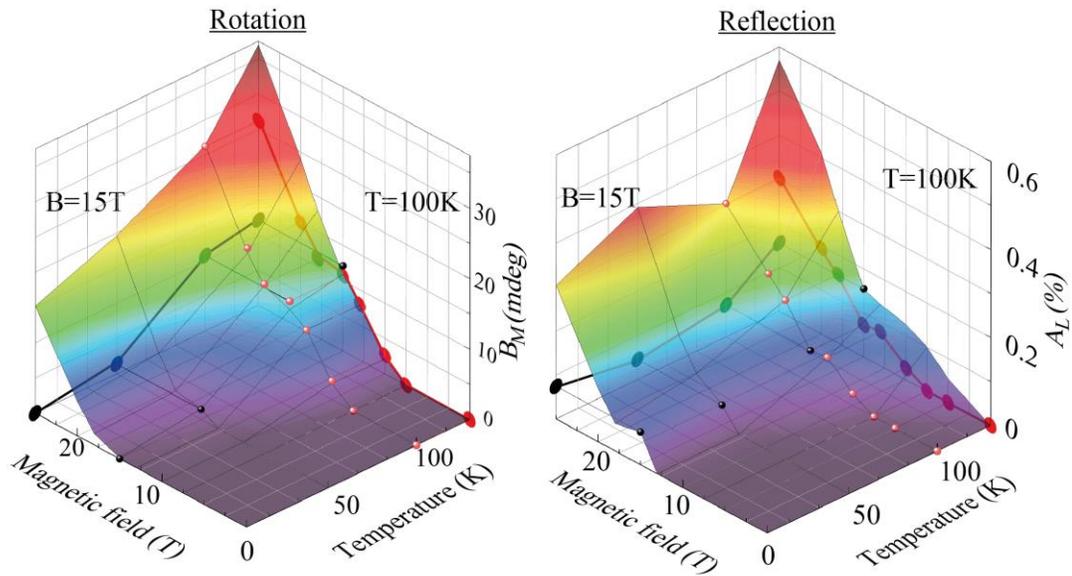

FIG. 9. Three-dimensional (3D) surface plots showing the amplitudes $A_L$ and $B_M$ of ultrafast magneto-structural changes. Plane curves correspond to a slice of the surface taken at the fixed temperature T = 100 K (red) or the fixed magnetic field H = 15 T (black). (a) The MOKE amplitude $B_M$ reveals ultrafast laser-induced magnetization. (b) The amplitude of ultrafast laser-induced reflectivity change $A_L$ is presumably assigned to the lattice expansion.

Hence if we assume that the dynamics of the reflectivity corresponds to the dynamics of the lattice expansion, this implies for sufficiently high magnetic fields the laser-induced magnetization emerges as fast as the lattice expands (see Fig. 7). Applying even higher fields does not accelerate either the lattice or the spin dynamics. It suggests that the lattice expansion and the magnetization emergence occur simultaneously on a sub-10 ps time scale defined by the thermal expansion of the lattice. Previous studies of the phase transition with the help of X-ray techniques reported similar characteristic times of the lattice expansion. The latter was shown to be defined by the speed of sound and the thickness of the film. Ref. [20] reported a shift of the Bragg peak already within the first 6 ps for a 47 nm thick film, while in Ref. [56] similar measurements on a film with a thickness of 100 nm showed a shift of the Bragg peak after 18.5 ps. XAS measurements on 30 nm thick demonstrated changes on a sub-10 ps time scale [18]. Thus, in the case of the film with a thickness of 40 nm, the lattice expansion should result in a reflectivity change on a sub-10 ps time scale as well.

The observed magneto-structural dynamic correlations suggest two possible scenarios. As the essence of the net magnetization induced in the ferromagnetic state is the angular momentum, conservation of the angular momentum must play in these scenarios the central role. In the first scenario, the effective antiferromagnetic exchange interaction acting between the two iron sublattices turns into a ferromagnetic one much faster than 2 ps, but the observed magnetization dynamics occurs on the scale of the lattice dynamics and is practically defined by the rate at which the lattice can exchange angular momentum with spins. Note that recent experiments revealed lattice dynamics upon ultrafast demagnetization of ferromagnetic Ni and Fe. The findings imply that the exchange of angular momentum between the lattice and spins at the sub-ps time scale is quite realistic [57,58]. Hence we rather favor the second scenario, where the effective field of the exchange interaction $\mathbf{H}_E$ changes on the scale of the lattice expansion, and the response of the spins as well as the exchange of angular momentum are fast enough to follow these changes.

## VII. SUMMARY

Exploring the peculiarities of the H-T diagram of FeRh, we identified possible spin arrangements in this material. Variation of magnetic fields and temperatures allows for obtaining three different types of laser-induced magnetization dynamics in FeRh.

These differences are explained by three different initial states of the spin order in FeRh, corresponding to collinear antiferromagnetic, canted antiferromagnetic, and ferromagnetic phases. It is shown that the applied magnetic field can accelerate the emergence of the ferromagnetic phase only to a certain limit. Moreover, we report on the time-resolved measurements of the reflectivity, which contain information about the dynamics of the structural changes in FeRh. We discovered a regime when the magnetization emergence coincides with the changes in reflectivity. It can mean that the spin dynamics is accelerated up to the time-scale of the accompanying structural changes.

We additionally support this interpretation by simulations showing that the observed changes in the magnetization dynamics are intrinsic even to the simplistic two-spin model and thus must be a general feature of all antiferromagnets and not only FeRh. The revealed dynamics agrees with the proposed interpretation that ultrafast heating of FeRh in the canted antiferromagnetic phase launches the instantaneous emergence of the magnetization with the rate of the lattice expansion. The results show that the magnetism-or-lattice causality dilemma is resolved by the simultaneous evolution of both actors.

**Acknowledgments**

We thank P. Albers, F. Janssen, S. Semin and Ch. Berkhout for technical support. The work was funded by the Nederlandse Organisatie voor Wetenschappelijk Onderzoek (NWO), HFML-RU/NWO member of the European Magnetic Field Laboratory (EMFL), the European Research Council ERC Grant Agreement No.101054664 273 (SPARTACUS). The authors would like to thank prof. F. Parmigiani for wide and helpful discussions. O.K. was supported by the Project of the Ministry of Science and Higher Education of the Russian Federation (Grant No. 13.1902.21.0006). The project has received funding from the European Union's Horizon 2020 research and innovation programme under the Marie Skłodowska-Curie grant agreement No 861300 (COMRAD). A. K. Z. acknowledges the financial support by the Russian Science Foundation, Project No. 22-12-00367.

**Author contributions**

A. V. K. formulated the idea of the project with inputs from P. C. M. C. and I. A. D.; R. M., S.K.K.P, and E.E.F. fabricated the samples. O.V.K. performed SQUID magnetometry with the participation of I.A.D.; I.A.D. performed magneto-optical characterization of the sample and measurements in low magnetic fields. I. A. D., K. H. P., G. L., T. G. H. B. performed the measurements in high magnetic fields. T. G. H. B., A. G. B., J. H. M., K. A. Z. and A. K. Z. developed the theoretical model. I. A. D. and A. V. K. wrote the manuscript with contributions from all co-authors. The project was supervised by P. C. M. C. and A. V. K.

**Competing interests**

The authors declare no competing interests.

**Data management**

The data can be accessed on request at https://doi.org/10.34973/3esy-hn91.

# Supplementary to "Ultrafast Emergence of Ferromagnetism in Antiferromagnetic FeRh in High Magnetic Fields"

I. A. Dolgikh*, T. G. H. Blank, A. G. Buzdakov, G. Li, K. H. Prabhakara, S. K. K. Patel, R. Medapalli, E. E. Fullerton, O. V. Koplak, J. H. Mentink, K. A. Zvezdin, A. K. Zvezdin, P. C. M. Christianen & A. V. Kimel

*Radboud University, Institute for Molecules and Materials, 6525 AJ Nijmegen, The Netherlands.*

*High Field Magnet Laboratory (HFML - EMFL), Radboud University, Toernooiveld 7, 6525 ED Nijmegen, The Netherlands.*

*Center for Memory and Recording Research, University of California, San Diego, La Jolla, California 92093-0401, USA.*

This supplementary information contains additional time-resolved measurements that complement the results and figures presented in the main text of the article. We also provide an extensive discussion of the detailed fitting procedure used. The sections are as follows:

- Time-resolved MOKE above room temperature;
- All time-resolved traces for the first 40 ps and up to 1.5 ns in high magnetic fields;
- Fitting procedure for the data obtained at low temperatures;

# I. Latency in time-resolved MOKE.

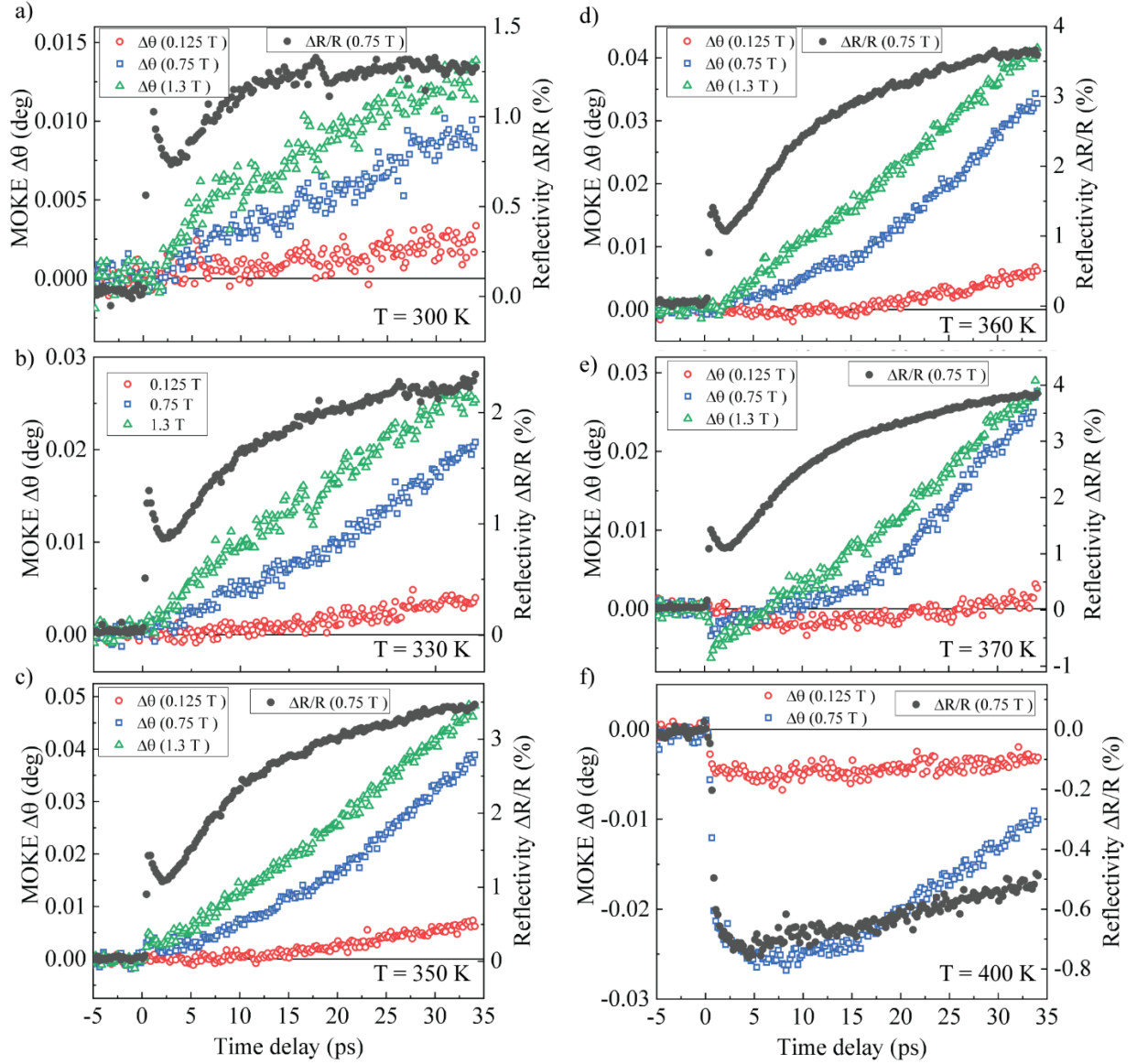

FIG. S1. Time-resolved polar MOKE and Reflectivity above room temperature (from 300 K to 400 K). The laser-induced reflectivity change (black filled circles) and the polarization rotation as a result of the magneto-optical Kerr effect at applied magnetic fields of 0.125 T (circles), 0.75 T (squares), and 1.3 T (triangles).

In preparation of the experiments described in the main article, we executed a time-resolved study of the AF to FM transition at room temperature in external magnetic fields up to 1.3 T. The same procedure as discussed in the manuscript was followed. We probed simultaneously both parts of the laser-induced response: P-MOKE and Reflectivity being the differential and summary signals of the p- and s- polarization components respectively.

It has been widely accepted that on a time-scale longer than 2 ps, the MOKE is a reliable probe of the magnetization dynamics [1] while the reflectivity change in FeRh is due to both the dynamics of the electrons and the lattice. In particular, it is well accepted that a rapid increase and a partial recovery of the transient reflectivity $\Delta R/R$ on the scale up to 1-2 ps is due to the electron temperature dynamics [2] and the slower dynamics of $\Delta R/R$ with a characteristic time of $\tau_\text{L} = 4 - 5$ ps can be attributed to the lattice expansion [2,3]. Although there is a slight discrepancy between $\tau_\text{L}$ estimated for the dynamics below 200 K and at 300 K, one must keep in mind that the sound velocity in compounds with structural transitions is not constant as a function of temperature. In particular, heating the sample towards the transition temperature can result in a decrease of the sound velocity and its non-linear behaviour with a jump at $T_\text{c}$, which was observed in materials with a metamagnetic transition from the AF to the FM phase [4,5].

Figure S1 shows the laser-induced dynamics of both the P-MOKE and reflectivity measured in FeRh at room temperature for fields of 0.125 T, 0.75 T, and 1.3 T applied perpendicular to the sample plane. Details of the experimental setup are explained in the main section. The 800 nm pulse of 4.1 mJ/cm$^2$ with an excitation spot size of 50 μm in diameter does not lead to sub-picosecond dynamics of the net magnetization. A study of time-domain THz emission spectroscopy on a similar sample showed that the sub-picosecond magnetization dynamics is due to a pump-induced expansion or demagnetization of pre-existing FM domains and not due to the phase transition from the AF to the FM phase [6]. Magnetization probed with the help of the P-MOKE emerges in the medium on a much longer time-scale in agreement with Refs. [7,8] following route type 1 from Fig. 1. For all traces from Fig. S1 dynamics occurs with a latency similar to what was observed at 5 T and 200 K, where latency in the emergent magnetization is present during the first 18 ps (see Fig. 5). Here, its duration decreases from 15 ps to 1.6 ps in 1.3 T. The observed rise in the characteristic time can be connected with the significant temperature difference. Investigating even lower temperatures, e.g. 100 K in magnetic fields up to 25 T which were obtained at the High Field Magnet Laboratory in Nijmegen (Fig. S2), we could not observe any latency. However, according to our estimation of the temperature rise provided by the pump, we were not heating enough to induce ferromagnetism at 5 T. All time-resolved dynamics above 7.5 T occurred according to Route type 2.

All traces of Fig. S2 were fitted using the same fit functions as in Fig. 7 of the main article. The details of the fits are described in Section IV of this Supplementary information.

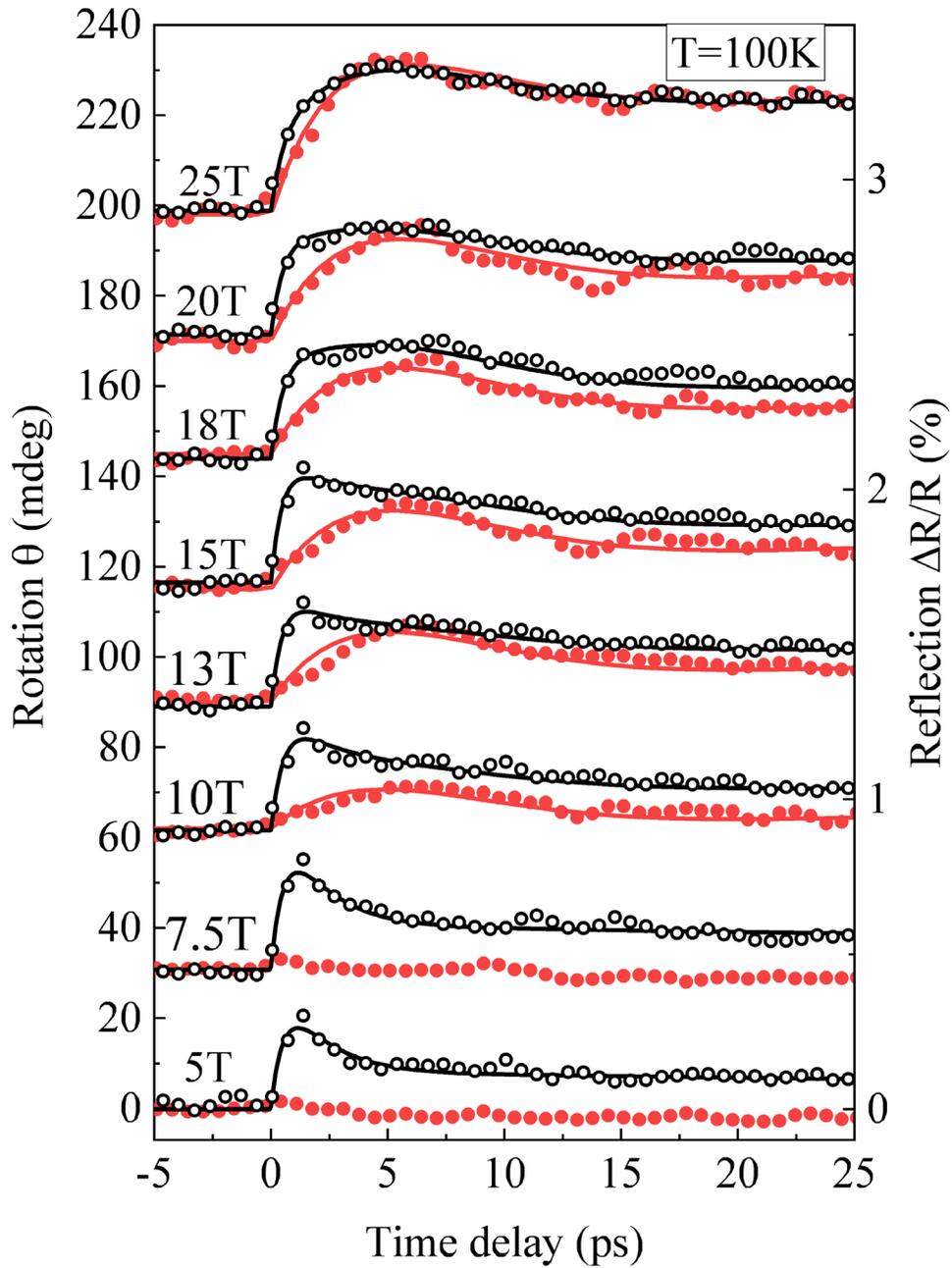

FIG. S2. Ultrafast kinetics of the magneto-structural transition. (a) The polarization rotation induced by the magneto-optical Kerr effect (red) and reflectivity change $\Delta R/R$ (black) at 100 K and various magnetic fields. The open circles represent the experimental data and the solid lines are their respective fits. The curves are plotted with an offset along the y-axis.

## II. Dynamics of the laser-induced MOKE used for Fig. 7

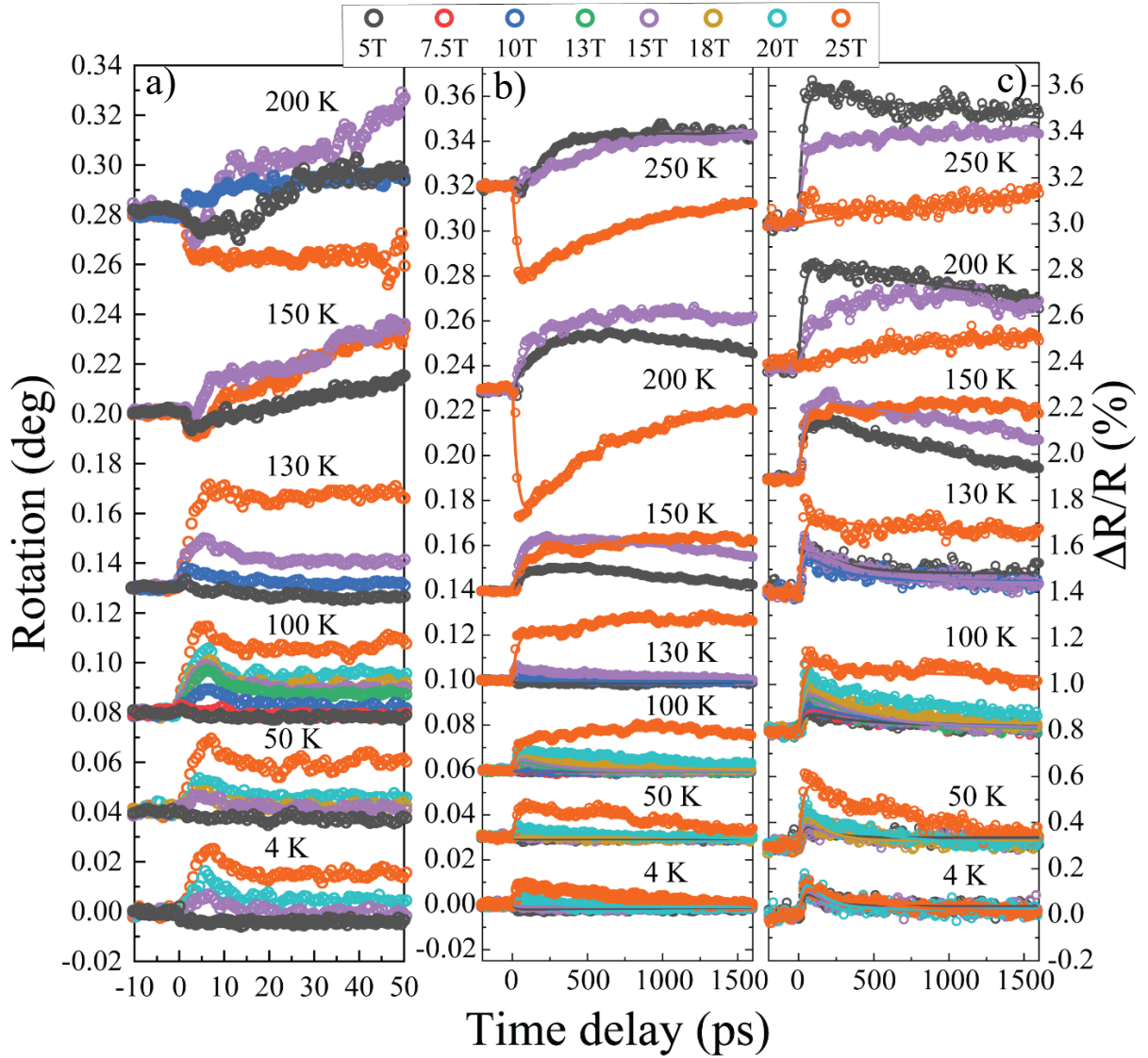

FIG. S3. Time-resolved Magneto-optical Kerr effect. Panel (a) shows time-resolved MOKE signals for time delays up to 50 ps. This data was used for building the phase diagram shown in Fig. 2 of the main article. Panel (b) shows laser-induced changes in the MOKE signal for delays up to 1.5 ns; Panel (c) shows the changes in reflectivity. The data was obtained in the range of temperatures from 4 K to 250 K. The external magnetic field was varied from 5 T to 25 T. Different colors correspond to different magnetic fields, denoted at the top of the figure.

Figure S3(a) shows the dynamics of the transient polarization rotation of the probe pulse obtained in magnetic fields up to 25 T during the first 50 ps after excitation, and Fig. S3(b,c) shows MOKE together with the reflectivity during 1.5 ns after pumping. Both measurements were performed simultaneously using the pump-probe method. The observed dynamics is very sensitive to both the applied magnetic field and the temperature. For instance, at 4 K and for magnetic fields below 20 T we hardly observe any laser-induced polarization rotation. The dynamics of the transient reflectivity measured in fields below 20 T are nearly independent on the field and have a form typical for the case when only an electronic contribution to the transient reflectivity is expected [1]. An increase of the field results in both an increase of the laser-induced polarization rotation and an additional contribution to the transient reflectivity. At higher temperatures, the field influences the laser-induced dynamics in the polarization rotation, and the reflectivity becomes even higher. In the temperature range of 4 K to 150 K, both the laser-induced polarization rotation (probing the net magnetization) and the reflectivity (probing the lattice expansion) increase when approaching the phase transition (Fig. S1). At temperatures above 150 K and fields above 20 T, we observe an appreciable change of the laser-induced polarization rotation where the amplitude changes sign. Such a change of sign occurs because the medium is already in the state with co-existing AF and FM phases under these experimental conditions. Ultrafast laser excitation results not only in the transition from the AF to the FM phase but also in the demagnetization of the co-existing FM phase. If the FM phase dominates over the AF phase state, the signal originating from the laser-induced demagnetization will dominate because the sample resides uniformly in the FM phase.

## III. Comparison of the amplitudes of MOKE and Reflectivity

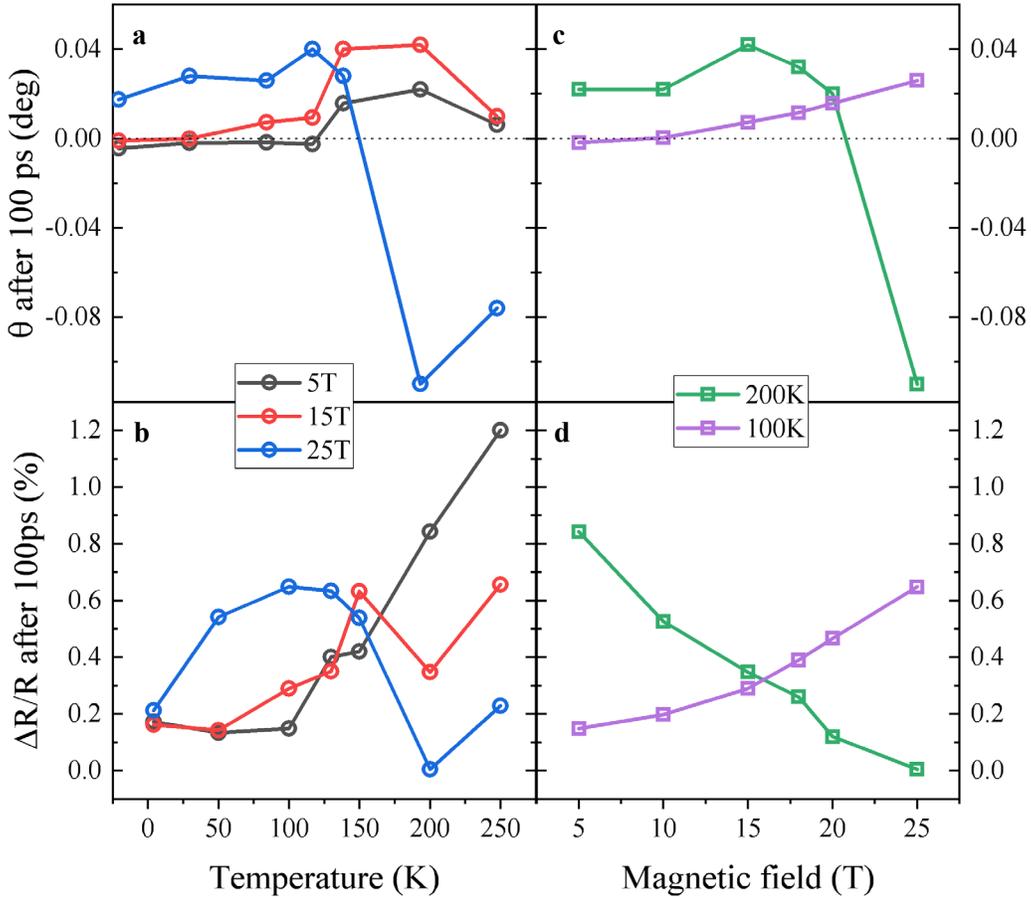

FIG. S4. The laser-induced polarization rotation and reflectivity change at the time delay of 100 ps extracted from Fig. S2. Panels a and c correspond to the polarization rotation at fixed fields and temperatures, respectively; Panels b and d correspond to the reflectivity at fixed fields and temperatures, respectively.

Figure S4(a,c) shows the MOKE signal strength at the time delay of 100 ps after the ultrafast laser excitation as a function of temperature at fixed fields (Fig. S4(a)) and as a function of the field at fixed temperatures (Fig. S4(c)). It is seen that up to 150 K, the MOKE only increases upon the field or temperature increase. The reflectivity signal also shows a similar trend (Fig. S4(b)). This observation once again confirms that, at least for the measurements up to 150 K, both the transient MOKE and the reflectivity contain information about the ns kinetics of the phase

transition from the AF to the FM state. The observed trends measured at 100 ps are in excellent agreement with the trend at the sub-10 ps time scale. Further increasing the magnetic field and the temperature leads to the change of a sign of θ, which indicates the transition to the FM phase. Besides this, we should take into account the time difference with the process of the phase transition. The response from the FM nucleus is on a sub-picosecond time scale as can be seen from Figure S3. Therefore, Fig. S4 does not leave any further doubts that the dynamics shown in Fig. 7 of the main article reveals the ultrafast kinetics of the phase transition from the AF to the FM state.

The dependencies shown in Fig. S3 can be fitted with two exponential contributions corresponding to growth with a characteristic time of roughly 0.2 - 0.5 ns and a relaxation with a decay time of 2 ns. The time scale of the growth is in agreement with the data reported earlier and must be due to the expansion process of the FM phase [1,8,9]. The characteristic time of the growth weakly depends on the applied magnetic field. Its trend is opposite to the one expected if the growth is due to an alignment of the magnetization of the FM domains. In particular, the characteristic time measured at 250 K and 15 T ($\tau \approx 600$ ps) is even slower than the one measured at 5 T ($\tau \approx 250$ ps). Therefore, it shows that the growth mechanism cannot have a magnetic origin and must be assigned to thermalization and the establishment of thermal equilibrium between phonons and spins. The relaxation corresponds to the time scale of cooling down.

## IV. Fitting procedure of the time-resolved MOKE traces below 150 K.

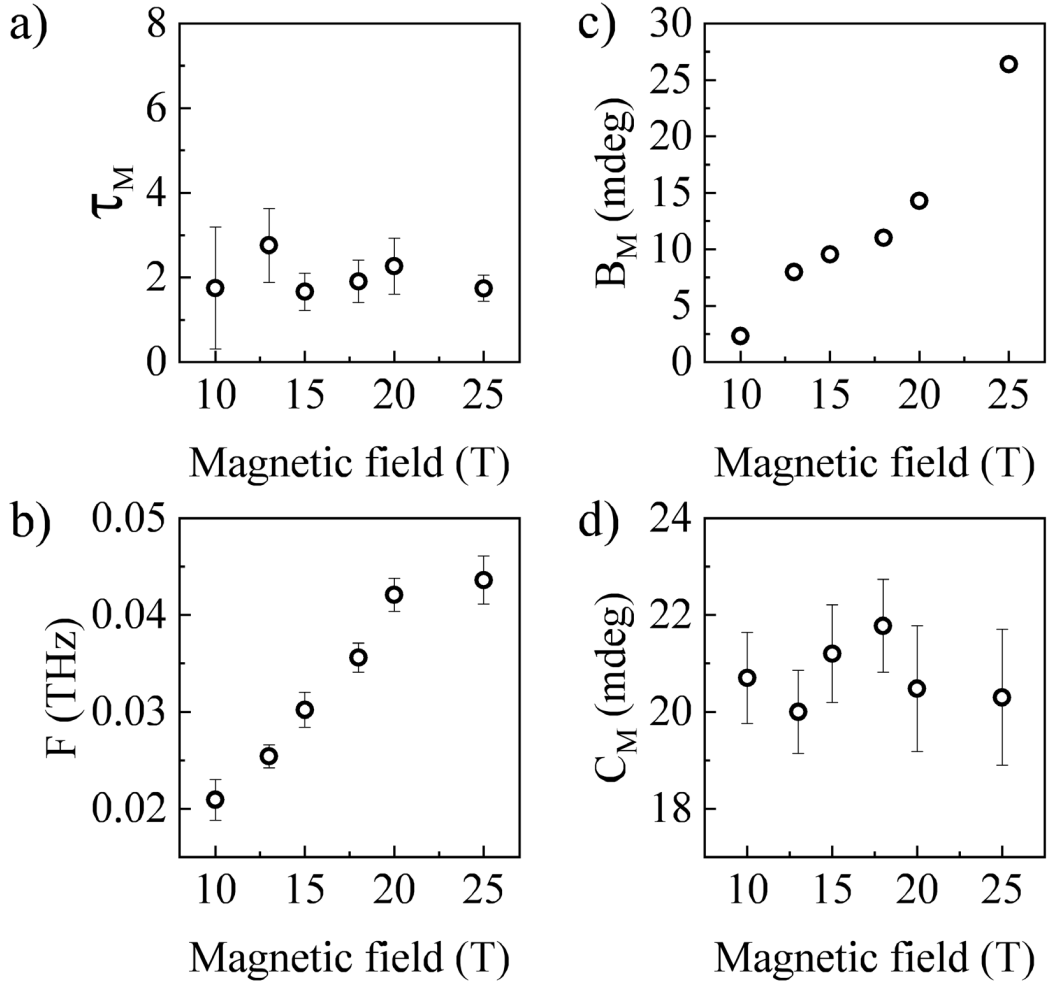

FIG. S5. Fitting parameters of the laser-induced MOKE signal at $T = 100$ K. Magnetic field dependence of the characteristic rise time of the FM nucleation (a), oscillation frequency (b), and the amplitudes of the exponential rise (c) and the oscillation (d).

In Figure 7 of the main article and in Figure S2, the solid lines correspond to fits that have been obtained using a more extended fitting function than the simple exponential rise used for the traces obtained at high temperatures. As was mentioned in the article, the transients exhibit an extra oscillatory behavior in addition to the exponential rise. To account for this part of the signal we used the following function to fit the MOKE signal:

$$\Delta\theta(t) = B_M \left(1 - e^{-\frac{t-\Delta\tau}{\tau_M}}\right) + C_M \sin(2\pi F\, t\,)e^{-\gamma t}. \tag{1}$$

The exponential rise, saturating at a certain level from 10 ps onwards, is primarily governed by the first term. The second term captures the presence of a heavily damped oscillation that emerges immediately after the laser excitation. $B_M$ and $C_M$ are the amplitudes of the exponential rise and the oscillation respectively, $\tau_M$ – is the characteristic rise time. The latency $\Delta\tau$ is present only below 5 T and is not observed together with the oscillation. Thus, in the following procedure we take $\Delta\tau = 0$. The oscillation is strongly damped: for all the transients we observe less than one period, so we fixed the decay rate at $\gamma = 0.15\, \frac{1}{ps}$. The frequency $F$ was left as a free parameter in the first iteration of the fitting. In the magnetic field dependence depicted in Figure S5a, starting from 10 T, we observe that the time constant remains unaffected by the field strength. This enables us to calculate an average time constant by considering measurements obtained at various external magnetic fields, which results in a value of $\tau_M = 2.0 \pm 0.4$ ps, as described in the text. Hence, for the fits shown in Fig.7 and in the rest of the paper, we fixed the time $\tau_M = 2$ ps and fitted only the amplitudes.

Subsequently, we examined the observed oscillation, which appears to exhibit a dependency on the magnetic field strength (Figure S5(b)). However, when we plot the residuals of the exponential fit (considering only the first term in Equation 1), we observe that the variations fall within the noise level (Figure S6). Therefore, the observation of only half of the oscillation period is insufficient to make a definitive judgment regarding its dependence on the magnetic field. For the sake of simplicity, we averaged it as well, getting $F = 33\pm9$ GHz, as used in the main manuscript.

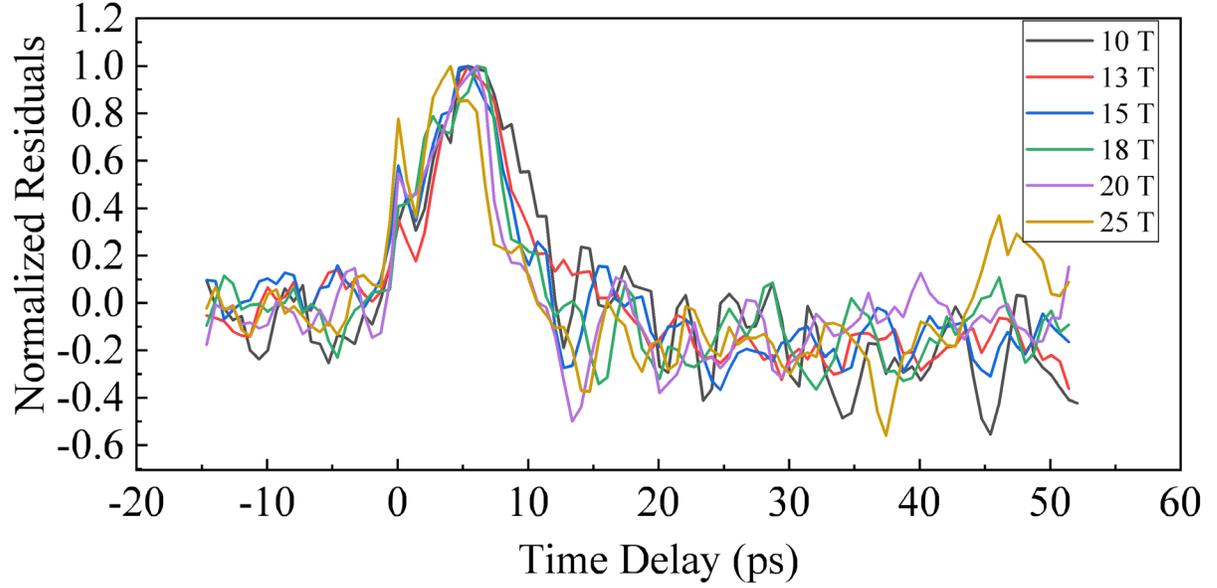

FIG. S6. The residuals were obtained by fitting the MOKE measurements with the exponential rise at T = 100 K. The transients representing a heavily damped oscillation introduced by the second term in Eq. 1. are shown for a range of external magnetic fields from 10 T to 25 T.

The reflectivity data is more complex and requires caution in its interpretation. This signal contains a fast response from the hot electrons and the relaxation of the heat to the lattice system. On the other hand, it does not contain information on magnetization and is not expected to show a dependency on the external magnetic field, assuming that no magnetoelastic phase transition occurs. Thus, we fitted the laser-induced reflectivity far away from the transition (at a low external field, when the MOKE signal does not show any positive contribution) with the function introduced in Ref [10], but with an additional scaling factor α:

$$\frac{\Delta R}{R}(t) = \left\{ A_{e1} * \left(1 - \exp^{-\frac{x}{\tau_{e1}}}\right) * \exp^{-\frac{x}{\tau_{e2}}} + A_{e2} * \left(1 - \left(\exp^{-\frac{x}{t_{e2}}}\right)\right) \right\} * \alpha * \exp^{-\frac{x}{t_{dif}}}. \quad (2)$$

This formula contains an electronic temperature increase in response to the laser excitation, which is represented by the first term with amplitude $A_{e1}$ and characteristic time $\tau_{e1}$.

| $A_{e1}$ | $A_{e2}$ | $\tau_{e1}$ | $\tau_{e2}$ |
|---|---|---|---|
| 0.79±0.04 % | 0.237±0.005 % | 0.71±0.10 ps | 1.65±0.15 ps |

Table 1. Parameters used for the electronic contribution in Eq. 2, which were obtained from fitting the laser-induced reflectivity in the absence of an external magnetic field at T = 100 K.

The second part describes temperature transfer to the lattice with amplitude $A_{e2}$ and time $\tau_{e2}$. The time $\tau_{e2}$ also defines the relaxation of the initial growth by modifying $A_{e1}$. Afterwards, the heat dissipates in the surroundings with $\tau_{dif}$.

To define the component responsible for the lattice expansion during the magnetoelastic transition, we fixed the parameters of the fit in Eq. 2, allowing only the change of the scaling factor $\alpha$. The fixed parameters are provided in Table 1. An example of fitting the laser-induced reflectivity without external magnetic field with Eq.2 is shown in Figure S7.

The scaling in the magnetic field reflects the changes in the electronic system due to the phase transition. The changes in the electronic band structure were reported at the subpicosecond time scale during laser-induced phase transition in a similar system [11]. The contribution is reduced in an external magnetic field and with the increase of the temperature, which is reflected in Figure S8(a). The time $\tau_{\text{dif}}$ shows a slight variation with temperature as well, but was fixed for the transients as a function of the applied magnetic field (Fig. S8(b)).

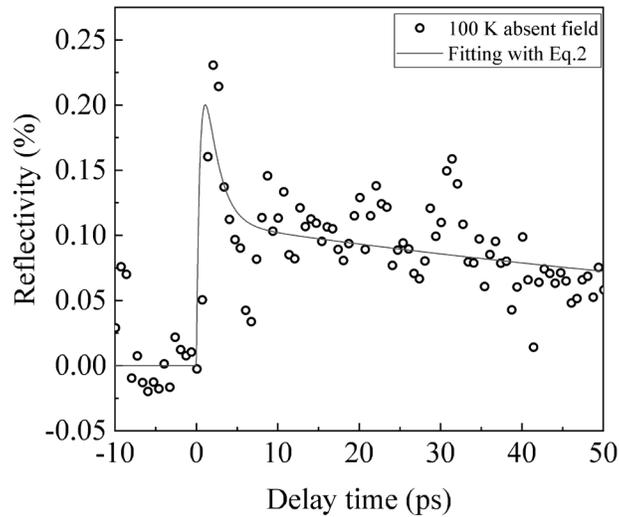

FIG. S7. Laser-induced reflectivity change in the absence of external magnetic fields at 100 K and the fit curve based on Equation 2 with the parameters shown in Table 1.

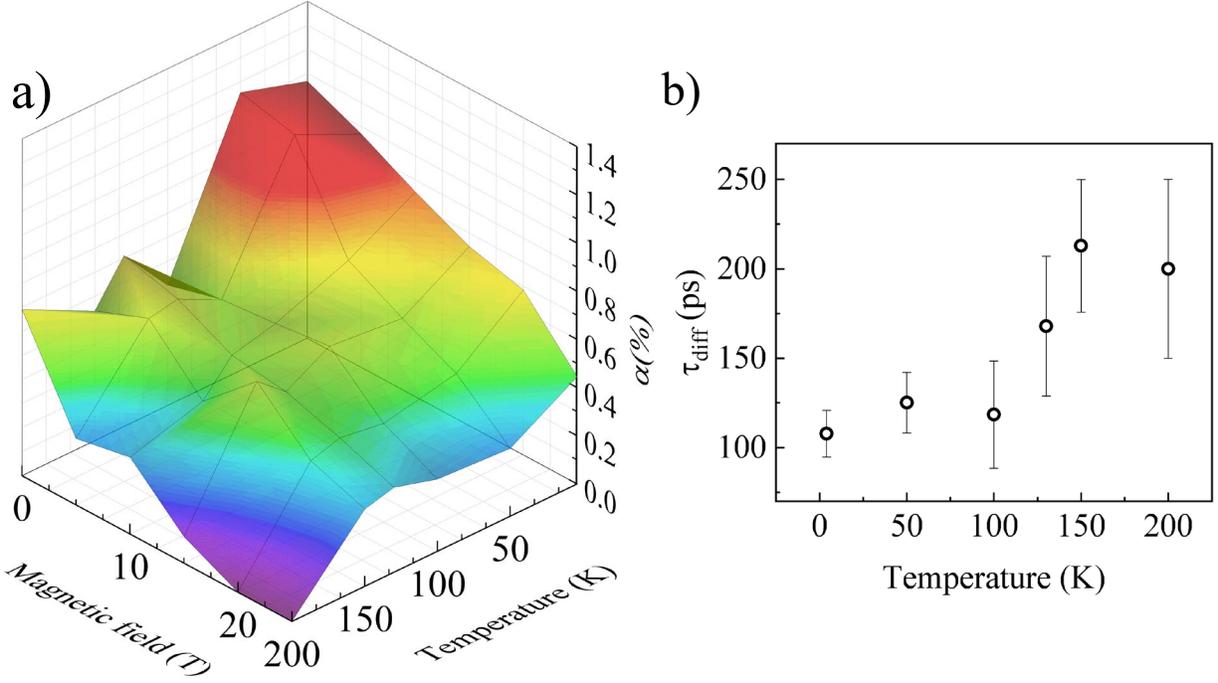

FIG. S8. Fitting parameters of laser-induced reflectivity represented by the Eq. 2. Magnetic field and temperature dependence of the scaling factor $\alpha$ (a), and the diffusion time (b).

Given that we now solely have a magnetic field-dependent response from the reflectivity, which can be attributed to the lattice expansion accompanying the phase transition, we proceeded to fit it using a similar function as employed for the MOKE signal, using the same decay rate.

$$\frac{\Delta R}{R}(t) = A_{\text{L}}\left(1 - e^{-\frac{t}{\tau_{\text{L}}}}\right) + C_R \sin(2\pi F\, t)e^{-\gamma t}, \qquad (3)$$

where $A_{\text{L}}$ and $\tau_{\text{L}}$ are the amplitude and the characteristic time presumably corresponding to the laser-induced lattice expansion, $C_R$ and $F$ are the amplitude and frequency of the harmonic oscillation.

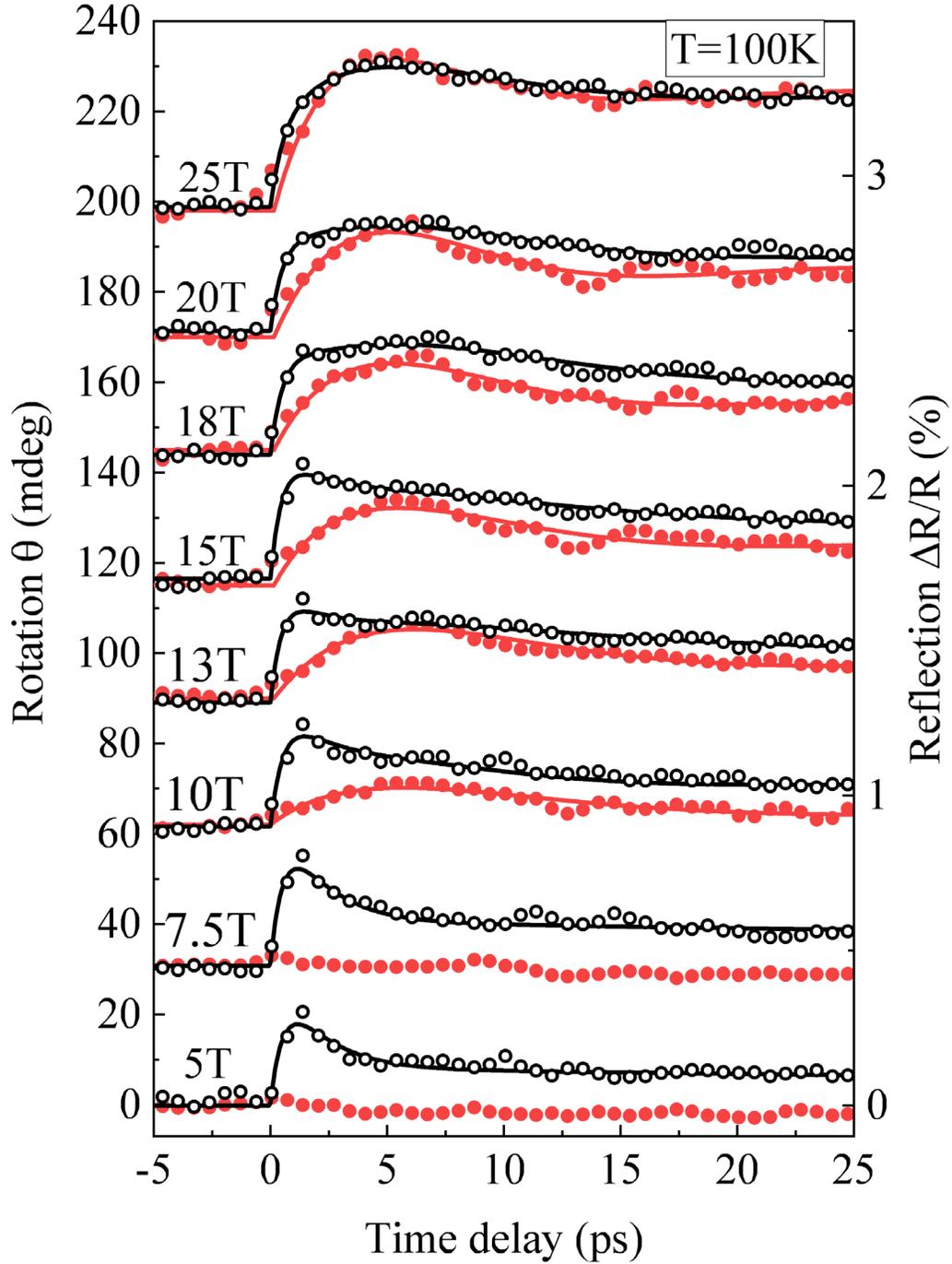

FIG. S9. The intermediate step of the fitting procedure used to obtain Fig. S2. The magnetic field dependence of the laser-induced MOKE and reflectivity in the external magnetic field at 100 K. The data is fitted with free time constant and frequency displayed in Fig. S10.

The results of the fit are shown in Figure S9 (black solid lines). The corresponding parameters are represented in Figure S10. From panel (a) it follows that the characteristic rise time $\tau_L$ does not depend on the magnetic field strength. The average value is $\tau_L = 2.7 \pm 0.9$ ps.

Taking into account the characteristic rise times for MOKE and reflectivity are close, we used $\tau = 2$ ps for both channels for the final fit (Figures 7 and S2).

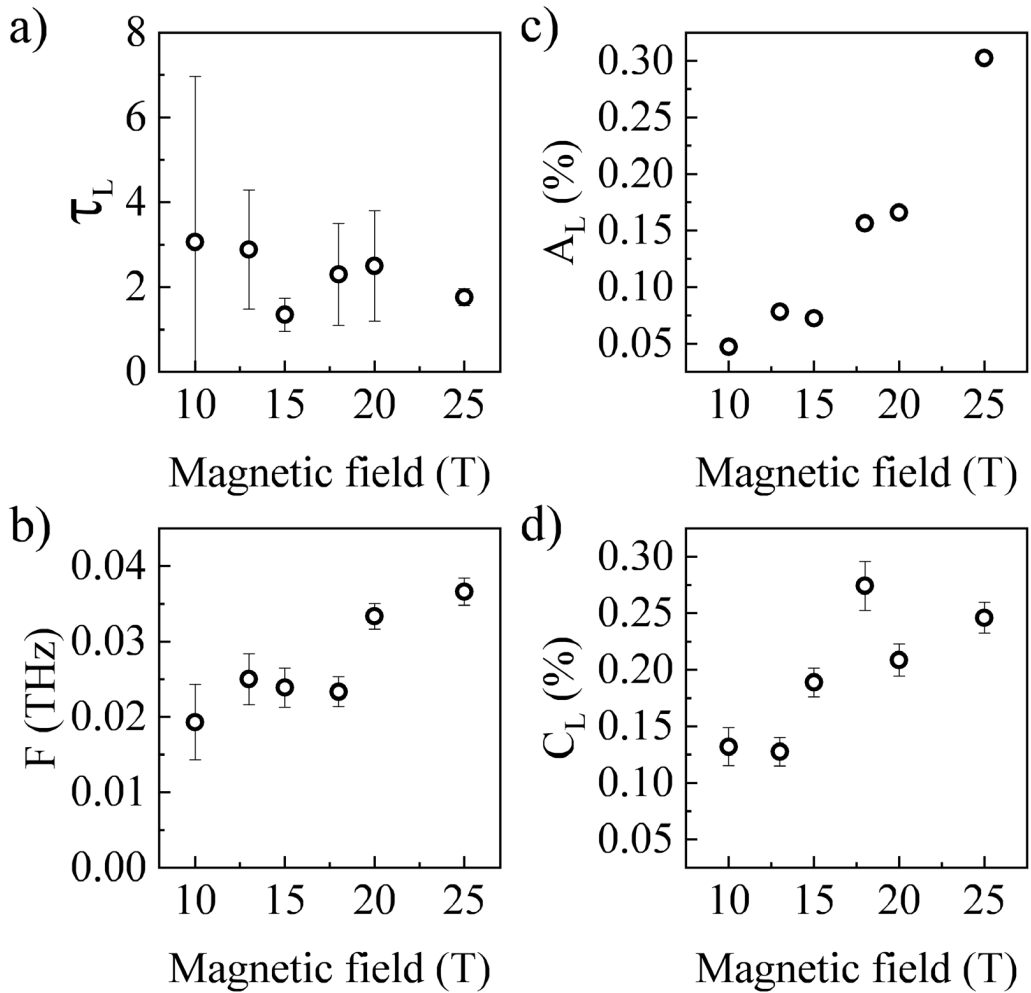

FIG. S10. Fitting parameters of laser-induced reflectivity at T = 100 K. Magnetic field dependence of the characteristic rise time of the slow exponential rise (a), oscillation frequency (b), and the amplitudes of the exponential rise (c) and oscillation (d).

Figure S10(b) displays the frequency values derived from fitting the slow component of the reflectivity, exhibiting a similar, but less pronounced trend to that shown in Figure S5(b).

By examining Figure S11, where residuals are plotted after fitting with only the first term in Equations 1 and 3, along with the electronic contribution described by Equation 2, it becomes evident that the damped oscillation is the same and present in both channels. When averaging the frequencies of the oscillation in reflectivity, we obtained F = 27 ± 7 GHz, which is close to the previously determined value of F = 33 ± 9 GHz for the MOKE transients. Thus, the frequency of the fit in Eq.3 can be also fixed to the value defined for the MOKE channel.

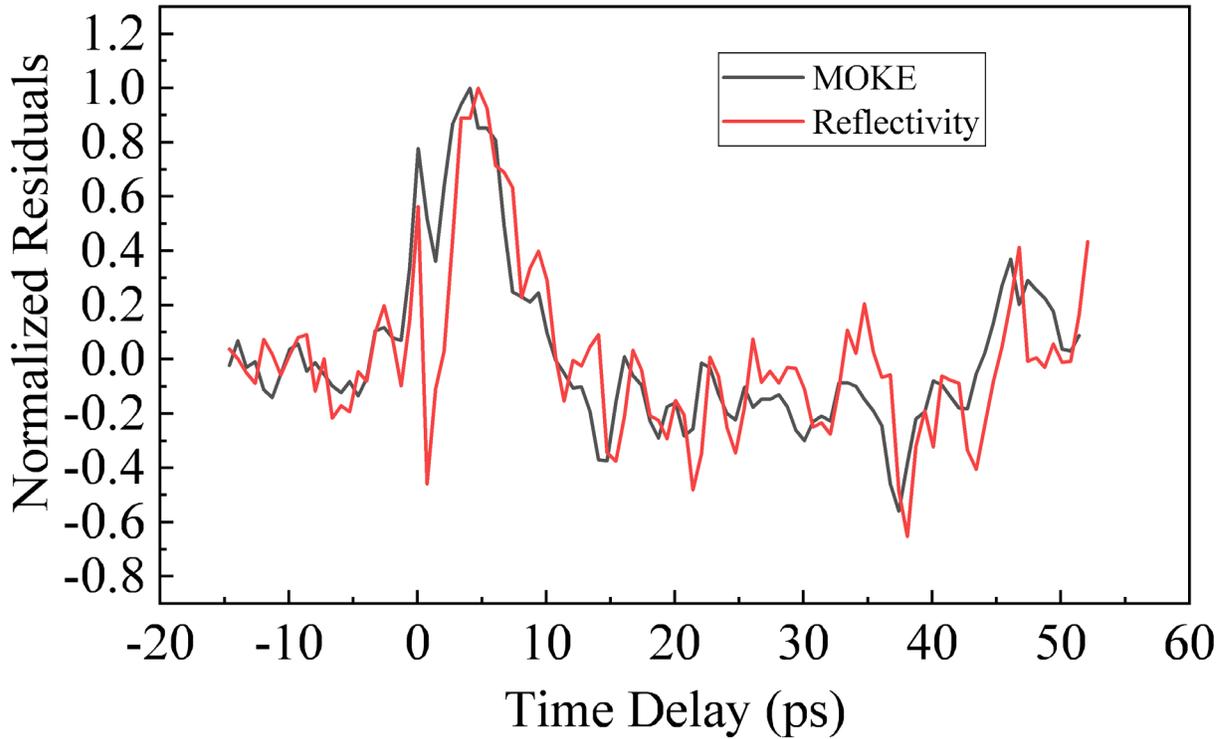

FIG. S11. The residuals obtained by fitting the MOKE measurements with the exponential rise, as well as the residuals after subtracting Equation 2 using the parameters from Table 1 and an exponential rise from the Reflectivity transient, are compared in the figure. The data were collected at a temperature of T = 100 K and a magnetic field strength of H = 25 T.

The resulting amplitudes are displayed in Figures S5(c,d) and S7(c,d). The final curves with those amplitudes and averaged time constant and frequency were used in the Figures 7 and S2.